\documentclass[iop,numberedappendix]{emulateapj}

\usepackage{graphicx}  
\usepackage{dcolumn}   
\usepackage{bm}        
\usepackage{amsfonts,amsmath,amssymb,mathrsfs}
\usepackage{color}

\usepackage{hyperref}
\hypersetup{
  colorlinks=true,        
  linkcolor=blue,         
  citecolor=cyan,         
}
\usepackage{times}

\def\vec#1{\mbox{\boldmath $#1$}}

\begin{document}

\shorttitle{Spatial Growth of CDI in Relativistic Rotating Jets} 
\shortauthors{Singh et al.}

\title{SPATIAL GROWTH OF CURRENT-DRIVEN INSTABILITY IN RELATIVISTIC ROTATING JETS AND THE SEARCH FOR MAGNETIC RECONNECTION}

\author{Chandra B. Singh\altaffilmark{1},
        Yosuke Mizuno\altaffilmark{2},
        and Elisabete M. de Gouveia Dal Pino\altaffilmark{1}}

\altaffiltext{1}{Department of Astronomy (IAG-USP),
                 University of S\~{a}o Paulo, S\~{a}o Paulo, Brazil;
                 csingh@iag.usp.br; dalpino@iag.usp.br}

\altaffiltext{2}{Institute for Theoretical Physics, Goethe University, 60438,
                 Frankfurt am Main, Germany; mizuno@th.physik.uni-frankfurt.de}

\begin{abstract}

Using the three-dimensional relativistic magnetohydrodynamic code \texttt{RAISHIN}, we investigated the influence of radial density profile on
 the spatial development of the current-driven kink instability along magnetized rotating, relativistic jets.
For the purpose of our study, we used a non-periodic computational box, the jet flow is initially established across the
computational grid, and a precessional perturbation at the inlet triggers the growth of the kink instability. We studied light 
as well as heavy jets with respect to the environment depending on the density profile. Different angular velocity amplitudes have been also tested.
The results show the propagation of a helically kinked structure along the jet and relatively stable configuration for the lighter jets. 
The jets appear to be collimated by the magnetic field and the flow is accelerated due to conversion of electromagnetic into kinetic energy. 
We also identify regions of high current density in filamentary current sheets,  indicative of magnetic reconnection, which are associated to
 the kink unstable regions and correlated to the decrease of the sigma parameter of the flow.
We discuss the implications of our findings for Poynting-flux dominated jets in connection with magnetic reconnection process. 
We find that fast magnetic reconnection may be driven by the kink-instability turbulence and govern the transformation of  magnetic 
 into kinetic energy thus  providing an efficient way to power and accelerate particles in AGN and gamma-ray-burst relativistic jets.
\end{abstract}

\keywords{galaxies: jets - instabilities - magnetohydrodynamics (MHD) - methods: numerical}

\section{Introduction}
Astrophysical relativistic jets are associated with several kinds of astrophysical systems which include X-ray binaries hosting stellar mass black holes
 (also denominated microquasars), active galactic nuclei (AGNs), and gamma-ray bursts (GRBs). 
According to their observed speed, one can separate them into: mildly relativistic ($v \sim 0.26c$) as in the case of microquasars like SS 433 and 
Cyg X-3 or low-luminous AGNs, highly relativistic ($v \sim 0.92c$) as in microquasars like GRS 1915+105 and GRO J1655-40,  extremely relativistic 
( $v \sim 0.99c$) as in some high-luminous  AGNs (blazars), and ultra-relativistic  jets ($v \sim 0.9999c$) as in the GRBs (e.g., Kato et al. 2008).
Observations of polarized non-thermal radiation (from radio to gamma-rays) indicate that these sources and their jets may be strongly magnetized
 particularly near the nuclear region. For instance, gamma-ray observations indicate a strongly polarized component related to the jet detected in 
the hard state of the microquasar Cyg X-1 (e.g., Laurent et al. 2011). 
Assuming equipartition between the magnetic energy and the kinetic power of the observed relativistic jets in X-ray binaries and magnetic flux conservation,
typical estimates of the magnetic fields in the innermost stable circular orbit of the stellar mass black holes (BHs)  are in the range $10^{7} - 10^{8}$ G 
(Piotrovich et al. 2014; see also de Gouveia Dal Pino \& Lazarian 2005;  Kadowaki et al. 2015; and Singh et al. 2015 for similar estimated values using
magnetohydrodynamical balancing in the inner accretion disk/BH region). 
In case of AGNs, physical parameters of the jets can be determined for instance from the analysis of the frequency dependence of the observed shift of the
core of relativistic jets using very-long-baseline interferometry (VLBI) radio data (e.g., from MOJAVE data base). Recent studies in this direction are
consistent with the extraction rotational energy from the BH in presence of a large scale magnetic field (e.g., Nokhrina et al. 2015). Also, observations
generally favour a cylindrical shape for the internal structure of  the outflow.
In another work by Marti-Vidal et al. (2015) regarding the AGN PKS 1830-211, the the detected polarization signal seems to be related to a strong 
magnetic field at the jet base. 
In the case of GRBs, if one relies only on the ambient magnetic field that is swept  (and amplified)  by the passage of the shock front of the jet head, 
one cannot explain the observed afterglow emission in these sources (Rocha da Silva et al. 2015). Nevertheless, the polarization studies associated with 
afterglow emission indicate the presence and survival of internally magnetized baryonic jets with large-scale uniform magnetic fields (Mundell et al. 2013; 
Wiersema et al. 2014). 

Several mechanisms of jet flow acceleration and collimation, namely, gas-dynamics acceleration, acceleration by radiation
and magnetohydrodynamic (MHD) mechanisms (Begelman et al. 1984 and references therein) have been proposed and it is also possible
that different mechanisms operate in different sources (de Gouveia Dal Pino 2005), or, otherwise, all mechanisms are operating simultaneously (Beskin 2010).
Currently there is ''democratic" consensus that the jets may arise from the combined effect of magnetic fields and rotation.
The influential works are the Blandford-Znajek (BZ) (Blandford \& Znajek 1977). and the Blandford-Payne (BP) model (Blandford \& Payne 1982; 
Bisnovatyi-Kogan \& Lovelace
2001 and references therein). The relativistic Poynting flux dominated (energy and angular momentum outflow carried predominantly
by the electromagnetic field) steady jets from the AGNs, microquasars and GRBs are believed to be driven by rotational energy
of the black holes as invoked in the BZ model. Whereas the quasi-relativistic matter dominated steady jets are driven by rotational
energy of accretion flow due to magneto-centrifugal mechanism as in the BP model. However there can be other alternative mechanisms
such as gradient of magnetic and gas pressure. If the jet has sufficiently large specific enthalpy and is overpressured, 
the relativistic jets can be powerfully boosted by the propagation of rarefaction wave from the interface between jet and ambient medium 
(e.g., Aloy \& Rezzolla 2006; Mizuno et al. 2008). This acceleration mechanism has been produced multiple recollimation shock/rarefaction structure 
in the relativistic jets which related stationary features in observed relativistic jets (e.g., Mizuno et al. 2015)
 
General relativistic MHD (GRMHD) simulations with accretion
disks around spinning black holes have shown the presence of high Lorentz factor Poynting dominated spine, mildly relativistic
matter dominated sheath and sub-relativistic wind as components of outflows and jets (Abramowicz \& Fragile 2013; Yuan \& Narayan
2014 and references therein). The spine-sheath structure of the jets have been observed in some nearby FRI radio galaxies, like
M87 (Kovalev et al. 2007) and 3C84 (Nagai et al. 2014),  FRII radiogalaxy like Cyg A (Boccardi et al. 2015) and blazars like
Mrk501 (Giroletti et al. 2004) and 3C273 (Lobanov \& Zensus 2001) using high-resolution VLBI technique. It is believed that the 
relativistic jet should be kinetically dominated at the large scales though it can be Poynting flux dominated at its base (Sikora et al. 2005; 
Giannios \& Spruit 2006). On one hand, it has been suggested that the conversion of magnetic energy to kinetic energy can involve gradual
 acceleration due to adiabatic expansion of the outflow and the observed emission can be caused by dissipation of field (if the magnetic field 
has the right geometry and scale) as result of magnetic reconnection due to current-driven kink instability and /or other instabilities 
such as the Kruskal-Schwarzschild instability in a striped wind (Granot et al. 2015 and references therein). Other models suggest that
 the acceleration efficiency can be increased by relaxing the assumption of axisymmetry leading to non-axisymmetric instabilities that can
 randomize the magnetic field orientation leading to efficient acceleration similar to thermal acceleration of relativistic outflows.
 Recent advances in numerical solutions for relativistic MHD have led to significant progress to understand more about the magnetic jet launching,
propagation and acceleration (Kumar \& Zhang 2015 and references therein), however the stabilities properties of jets are  still under debate.

Important signatures of the jet morphology are given by the non-thermal radiation that they emit from radio to gamma-rays. This is produced by
 relativistic particles accelerated along the flow. In regions where shock discontinuities are evidenced like in the bright internal knots or at 
the terminal bow shock at the jet head, the particle acceleration is attributed to diffusive Fermi shock acceleration. However, in magnetically 
dominated regions specially near the nuclear region, shocks are faint and particle acceleration is attributed to Fermi process in magnetic 
reconnection discontinuities (de Gouveia Dal Pino \& Lazarian 2005; Kowal et al. 2011, 2012; Giannios 2010; Sironi \& Spitkovsky 2014; de Gouveia Dal Pino \& Kowal 2015).

When jets are magnetically dominated, they are likely to experience current driven instability (CDI) which in turn may drive reconnection, while in case of kinetically
dominated jets, Kelvin-Helmholtz instability (KHI) driven by velocity shear between the jet and the surrounding medium comes into play.
Using VLBI, twisted structures are observed in several AGN jets on sub-parsec, parsec and kilo-parsec scales (Gomez et al. 2001;
Lovanov \& Zensus 2001; Lovanov, Hardee \& Eilek 2003).  These structures can be formed due to the jet precession and/or MHD 
instabilities like CDI or KHI. Jet structures attributed to instabilities are even seen in laboratory experiments (Hsu \& Bellan 2002, 2005). 
Nonrelativistic and relativistic magnetic jet simulations have confirmed that the helical twisted structures in the jets are due to CDI or KHI 
(Hardee 2013; Mizuno et al. 2014 and references therein). 
Twisted structures and the sense of twist can be triggered by the precession of the jet base (Begelman et al. 1980) or by rotation of the jet 
fluid or by magnetic field helicity. In MHD models for jet collimation and acceleration, a toroidal magnetic field ($B_{\phi}$) is wound up due to rotating disk and/or
spinning black hole or neutron star and eventually dominates over the poloidal magnetic field ($B_{p}$) because $B_{p}$ falls off
faster with expansion and distance. The strong toroidal magnetic field is CD kink mode unstable (Begelman 1998). According to 
Kruskal-Shafranov (KS) instability criterion, cylindrical MHD configurations in which $B_{\phi}$ dominates are
violently unstable to the m=1 kink mode (screw instability). The KS criterion for instability indicates that the 
instability develops if the length of a static plasma column is long enough for the field lines to go around the column at least once
i.e. $|B_{\phi}/ B_{p}| > 2 \pi R/z$ (e.g. Bateman 1978) where $R$ is the cylindrical radius and z  the length of the jet.
Astrophysical jets can be sometimes violently unstable and eventually the power of the BZ mechanism
may be significantly suppressed (Li 2000). Non-relativistic analytical works (Appl et al. 2000; Baty 2005) and numerical
simulations (Lery et al.2000; Baty \& Keppens 2002) show that CDI growth rates exceed KHI growth rates for sufficiently large magnetic
field strength and helicity. CDI mode becomes dominant at increased helicity as the instability results from an imbalance between
destabilizing toroidal and stabilizing poloidal magnetic fields.
One approach to study simplistic picture of relativistic jet is to consider force-free approximation in which only the charges,
currents and fields are considered while the inertia and pressure of the plasma are ignored. This is valid whenever Poynting
flux dominated jets are taken into account. Cylindrical force-free jets are kink stable if the  axial magnetic field, $B_{z}$, is independent of R 
(Istomin \& Pariev 1994, 1996) but are kink unstable if $B_{z}$ decreases with increasing distance from the axis
(Begelman 1998; Lyubarsky 1999). 
The classical KS criterion can be extended if the stabilizing effect of the relativistic field
line rotation is taken into account which ensures the importance of the BZ mechanism at large distances from black hole
(Tomimatsu et al. 2001). The jets are unstable only if both the KS criterion and the condition: $|B_{\phi}/ B_{p}|  > R |\Omega_{F}|/c$, 
where $\Omega_{F}$ denotes the angular velocity of the field lines and $c$ the speed of light, are satisfied. All of the above mentioned
analyses have been done in static reference frame with the jet of infinite extent. Further, assuming a rigid wall enclosing the jet,
a class of force-free cylindrical jet equilibria including the important effects of the poloidal field curvature was studied and
the kink mode growth was found to be much slower than that expected from the KS instability criterion (Narayan et al. 2009).

The temporal development of the CD kink instability in relativistic cylindrical jets using periodic computational box has been
explored by several groups. Using the 3D general relativistic magnetohydrodynamical (GRMHD) code \texttt{RAISHIN}, 
Mizuno et al. (2009) studied the instability of a helically magnetized relativistic non-rotating force-free static plasma column 
and showed that the initial configuration is strongly distorted
but not disrupted by the kink instability. The growth rate and nonlinear evolution of the CD kink instability depends
moderately on the density profile and strongly on the magnetic pitch profile. This work was further extended by 
Mizuno et al. (2011) where the influence of a velocity shear surface on the development of the CD kink instability in a sub-Alfv\'{e}nic
non-rotating jet was investigated. It was found that the helically distorted density structure propagated along the jet with speed and flow
structure dependent on the ratio between the radius of the velocity shear surface and the characteristic radius of the helically twisted
force-free magnetic field. Adopting a similar approach as that of Mizuno et al. (2009), O’Neill et al. (2012) studied local
models of comoving magnetized plasma columns in force-free, pressure- and rotation-supported equilibrium configurations using the 
3D relativistic MHD code \texttt{ATHENA} (Beckwith \& Stone 2011). They found the development of dissimilar structures in different configurations,
however all those configurations of plasma columns were nominally unstable to CDI. In another interesting work of
uniform density, small magnetic pitch and highly magnetized plasma column, Anjiri et al. (2014) studied and quantified the
growth of CDI and KHI considering  three different cases corresponding to static, trans-Alfv\'{e}nic and super-Alfv\'{e}nic non-rotating jets using the MHD module
of  the \texttt{PLUTO} code (Mignone et al. 2007). In Mizuno et al. (2012), the influence of jet rotation and differential motion on the CD kink
instability was studied for rotating helically magnetized relativistic jet with decreasing density profile and it was found that in
accordance with the linear stability theory, the development of the instability depends on the lateral distribution of the poloidal
magnetic field. If the poloidal field significantly decreases outward from the axis, then the initial small perturbations grow
strongly, and if multiple wavelengths are excited, then nonlinear interaction eventually disrupts the initial cylindrical 
configuration. When the profile of the poloidal field is shallow, the instability develops slowly and eventually saturates. 
Recently Porth \& Komissarov (2015) studied the causality and stability of non-rotating cylindrical static columns with purely
toroidal magnetic field and took into account the role of the jet expansion in atmospheres with power-law pressure distribution.
The initial configuration was perturbed in the same way as that of Mizuno et al. (2012) via the adding of the radial velocity component
to the jet velocity field. The simulations have been carried out using the \texttt{MPI-AMRVAC} code (Porth et al. 2014; Keppens et al. 2012) 
and the test configuration was found to be disruptively unstable to CD kink instability as compared to the force-free case.    

In order to take into account proper real conditions, some non-relativistic (e.g., Nakamura \& Meier 2004; Nakamura et al.
2007; Moll et al. 2008; Moll 2009) and relativistic (e.g., McKinney \& Blandford 2009; Mignone et al. 2010, 2013; Porth 2013)
MHD simulations of jet formation and propagation following the spatial properties of the development of the kink instability in the
jet have been performed. Using the 3D GRMHD \texttt{HARM} code (Gammie, McKinney \& Toth 2003), McKinney \& Blandford (2009) simulated 
rapidly rotating, accreting black holes producing jets containing both toroidal and poloidal field components and it was found
that despite strong non-axisymmetric disc turbulence, the jet propagated and accelerated without significant disruption or
dissipation with only mild substructure dominated by the kink mode. While Mignone et al. (2010) presented 3D simulations  of
relativistic magnetized jets carrying an initially toroidal magnetic field using the \texttt{PLUTO} code (Mignone et al. 2007) and found
that the toroidal field gives rise to strong CD kink instability leading to jet wiggling. Nevertheless,  it appears to be maintained
a highly relativistic spine along its full length. Furthermore, in another work by Mignone et al. (2013) it was concluded that jets
with moderate to high magnetization parameter and relatively small bulk Lorentz factor are most likely candidates for the
dynamical behavior observed in the Crab nebula jet. Porth (2013) studied the formation of relativistic jets from rotating
magnetospheres and the results showed a saturation of different mode perturbations including the kink mode before a notable
dissipation or even disruption was encountered.

Recently, Mizuno et al. (2014) studied the influence of the velocity shear and a radial density profile on the spatial development of the CD
kink instability along helically magnetized non-rotating relativistic cylindrical jets using a non-periodic computational box
and a precessional perturbation at the inlet triggering the growth of the kink instability. If the velocity shear radius is located
inside the characteristic radius of the helical magnetic field, a  static non-propagating CD kink is excited as the perturbation propagates 
downstream the jet. On the other hand, if the velocity shear radius is outside the characteristic radius of the helical
magnetic field, the kink is advected with the flow and grows spatially down the jet. When the density increases with radius, 
the kink appears to saturate by the end of the simulation without apparent disruption of the helical twist which suggests
the relatively stable configuration for relativistic jets with tenuous spine and denser sheath. 

In this paper, we will investigate the influence of the jet rotation as well as density profile on the spatial development of the CD kink instability in a rotating
relativistic jet. Based on temporal studies, it is well known how the CD kink instability depends on the differential  motion, magnetic pitch profile 
and density profile (Mizuno et al. 2009, 2011, 2012).
For this study we will consider constant magnetic pitch and take the radius of the jet core to be the same as that of the characteristic radius of the helical magnetic
field. Basically, we will attempt to extend further the work of Mizuno et al. (2014) by taking into account different angular
velocity amplitudes of the jet and also compare with the temporal studies of rotating relativistic jets (Mizuno et al. 2012) 
by considering decreasing and increasing density profiles. 
We will also examine the potential development of magnetic reconnection sites along the jet triggered by the CD modes and its implications for 
magnetic energy dissipation which may be particularly important both in the determination of the transition of Poynting flux dominated to 
kinetically dominated regimes and for particle acceleration as stressed above, particularly in AGN and GRB jets (e.g., Zhang \& Yan 2011; 
Granot et al. 2015; de Gouveia Dal Pino \& Kowal 2015). 
 In fact, it is possible that the presence of turbulence in MHD flows can lead to fast magnetic reconnection (Lazarian \& Vishniac 1999).
 Zhang and Yan (2011), for instance, based on Lazarian and Vishniac (1999) and de Gouveia Dal Pino \& Lazarian (2005), proposed a model of GRB 
prompt emission where turbulence, magnetic reconnection and particle acceleration can be induced by internal collisions. In this work, we attempt 
to see such possible link between the turbulence induced by CD kink instability and magnetic reconnection.

In Section 2, the details of numerical method and setup are presented. Section 3 describes our numerical results and
final concluding remarks are in Section 4. 

\section{Numerical method and setup}
\label{sec:ns}

\begin{figure*}[ht!]
\begin{center}
\includegraphics[width=0.8\textwidth]{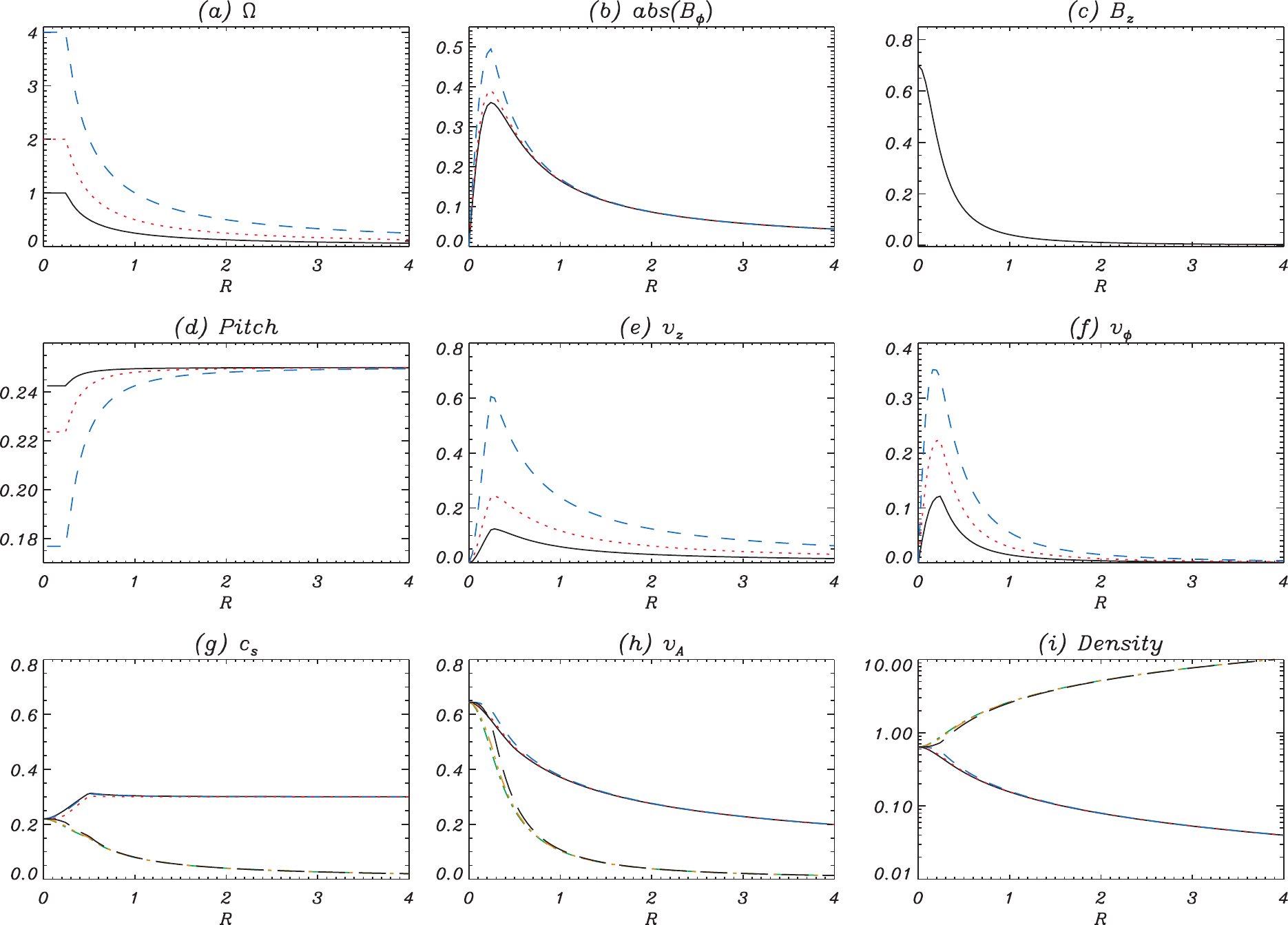}
 \end{center}  
\caption{Radial profiles of {\it(a)} the angular velocity $\Omega$ (given in units of $1/c$), {\it (b)} the toroidal magnetic field $|B_{\phi}|$ (in units of $\sqrt{4 \pi \rho_0 c^2}$), {\it (c)} the axial magnetic  field $B_{z}$ (also in units of $\sqrt{4 \pi \rho_0 c^2}$), {\it (d)} the magnetic pitch  $P= B_{\phi}/B_p$, {\it (e)} the axial velocity $v_{z}$ (in units of $c$), {\it (f)} the toroidal velocity $v_{\phi}$  (also in units of $c$), {\it (g)} the sound speed $c_{s}/c$,
{\it (h)} the Alfv\'{e}n speed $v_{A}/c$, and {\it (i)} the density $\rho$ (in units of $\rho_0$) for decreasing density profile with $\Omega_{0} = 1$ (black solid), $2$ (red dotted) and $4$ (blue dashed) and for increasing density profile with $\Omega_{0} = 1$ (green dash-dotted), $2$ (orange dash-two-dotted) and
 $4$ (black long dashed). }
\label{f1_1Dinit}
\end{figure*}

We perform special-relativistic magnetohydrodynamics (SRMHD) simulations using the three-dimensional general relativistic
 MHD code \texttt{RAISHIN} (Mizuno et al. 2006, 2011). The code setup for spatial development of the CD kink instability is similar
to the setup as mentioned in Mizuno et al. (2014). The computational domain is $8L \times 8L \times 30L$ in a cartesian (x, y, z) coordinate 
system with grid resolution of $\Delta L = L/30$ for the transverse x- and y-directions and $\Delta L = L/6$ for the axial z- 
direction. Outflow boundary conditions have been imposed on all surfaces except the inflow plane at $z=0$. A preexisting 
jet is established across the computational domain. 

A force free configuration is chosen for the initial configuration which is a reasonable choice for a Poynting flux dominated 
jet. Following previous works, the poloidal magnetic field component and the angular velocity of the jet are written in the following form

\begin{equation}
B_{z} = \frac {B_{0}}{[1+(R/a)^{2}]^{\alpha}},
\end{equation}

\begin{equation}
      \Omega=
\begin{cases}
      \Omega_{0}                    & \text{if } R \le R_{0}\\
      \Omega_{0} (R_{0}/R)^{\beta}  & \text{if } R > R_{0},
\end{cases}
\end{equation}

where $R$ is the radial distance in cylindrical coordinates, $B_{0}$ is the magnetic field amplitude, $\Omega_{0}$ is the 
angular velocity amplitude, $a$ is the characteristic radius of the magnetic field, $R_{0}$ is the radius of the core, $\alpha$ 
is the poloidal field profile parameter and $\beta$ is the angular velocity profile parameter. 

In the configuration chosen for the simulations, the initial poloidal and toroidal components of the drift velocity are given by,
\begin{equation}
v_{z} = - \frac{B_{\phi} B_{z}} {B^{2}} \Omega R,
\end{equation}

\begin{equation}
v_{\phi} = (1 - \frac{{B_{\phi}}^{2}} {B^{2}}) \Omega R.
\end{equation}

In the simulations, we choose $R_{0}$ = $a$ = $(1/4)L$, $\alpha$ = 1 so that the magnetic pitch  and the magnetic helicity are constant, and $\beta$ = 1.
The toroidal field component of the magnetic field is given by,

\begin{equation}
B_{\phi} = -\frac {B_{0}(R/a)[1 + (\Omega R/a)^{2}]^{1/2}} {[1 + (R/a)^{2}]}.
\end{equation}

A low gas pressure medium with pressure decreasing radially, similar to equation (2) and with $p_{0} = 0.02$ in units of $\rho_{0}c^{2}$
is considered (Mizuno et al. 2012) as follows:
\begin{equation}
      p=
\begin{cases}
      p_{0}                   & \text{if } R \le R_{p}\\
      p_{0} (R_{p}/R)^{\beta}  & \text{if } R > R_{p},
\end{cases}
\end{equation}
where $R_{p}$ is taken to be equal to L/2.

We note that this radial gas pressure profile favours  high Alfv\'{e}n  and low sound speeds in the system, which is compatible with the assumed initial 
force-free  magnetic field configuration where  the gas pressure gradient is not dominant. This behaviour will persist as the system evolves in time, 
as we will see below.

Two different radial density profiles: decreasing and increasing with radius are 
chosen. The decreasing density profile is given by $\rho = \rho_{1} \sqrt{B^{2}/B_{0}^{2}}$ and increasing density profile is given by 
$\rho = \rho_{1} \sqrt{B_{0}^{2}/B^{2}}$ where $\rho_{1} = 0.8 \rho_{0}$. The magnetic field amplitude is $B_{0} = 0.7$ in units of $\sqrt 
{4\pi\rho_{0}c^{2}}$ so that a low plasma $\beta$ near the axis is maintained.  
The equation of state is that of an ideal gas with $p = (\Gamma - 1) \rho e$, where $e$ is the internal specific energy density and the 
adiabatic index, $\Gamma = 5/3$ 
 \footnote{ We note that our choice of  adiabatic index is valid for a cold plasma where the sound speed is much smaller than the light speed. 
This condition is generally fulfilled in our simulations. Mizuno et al. (2009) tested different adiabatic indices and found no significant 
differences in  the growth of the CD kink instability among the models (see also Rocha da Silva et al. 2015 for further numerical tests on the 
dynamical behaviour of relativistic jets with different adiabatic indices.}

The specific enthalpy is $h \equiv 1 + e/c^{2} + p/\rho c^{2}$. The sound speed is given by 
$c_{s}/c \equiv (\Gamma p/\rho h)^{1/2}$, and the Alfv\'{e}n speed is given by $v_{A}/c \equiv [b^{2}/(\rho h + b^{2})]^{1/2}$, where 
$\vec{b}$ is the magnetic field measured in the comoving frame, $b^{2} = B^{2} / \gamma^{2} + ({\vec{v} \cdot \vec{B}})^{2}$ (Komissarov 
1997; Del Zanna et al. 2007), where $\gamma$ is the Lorentz factor of the jet.

To break the equilibrium state, a precessional perturbation is applied at the inflow using a transverse velocity component with a 
magnitude of $v_{\perp} = 0.01v_{j}= 0.009c$ and an angular frequency of $\omega_{j} = 2 \pi v_{j}/\lambda_{j}$ where $\lambda_{j} = 3L$ is a characteristic wavelength. 
In order to study the effect of jet rotation, simulations are performed with three different angular velocities, $\Omega_{0} = 
1$, $2$ and $4$. Radial profiles of the initial angular velocity, the magnetic field components, the magnetic pitch, the velocity components, the sound and 
Alfv\'{e}n speeds and the density for different angular velocity cases with $\alpha = 1$ are given in Figure \ref{f1_1Dinit}.

Here the initial angular velocity $\Omega_{0}$ and radial density profiles of the rotating, relativistic jets are such that the
flow is kept sub-Alfv\'{e}nic inside the core for both decreasing density cases (D1, D2 and D4) with $\Omega_{0} = 1$, $2$ and $4$,
 and increasing density cases (I1, I2 and I4)  with $\Omega_{0}= 1$, $2$ and $4$. This will allow us to focus on the development of the 
CD kink instability. The different initial parameters for the simulated models are given in Table \ref{table1}.

\begin{table}
\begin{center}
\begin{tabular}{lcc}
\hline
\hline
{Case} & {$\Omega_{0}$} & {Density profile} \\
\hline
\texttt{D1} & $1$ & {Decreasing} \\
\texttt{D2} & $2$ & {Decreasing} \\ 
\texttt{D4} & $4$ & {Decreasing} \\
\texttt{I1} & $1$ & {Increasing} \\
\texttt{I2} & $2$ & {Increasing} \\ 
\texttt{I4} & $4$ & {Increasing} \\
\hline
\end{tabular}
\caption{Basic properties of the various cases simulated. Listed are the initial angular velocity $\Omega_{0}$ and initial density profile of jets.}
\label{table1}
\end{center}
\end{table}

\section{Simulation results}
\label{sec:res}

\subsection{Development of the kink instability}

\begin{figure*}[!ht]
\begin{center}
\includegraphics[width=0.9\textwidth]{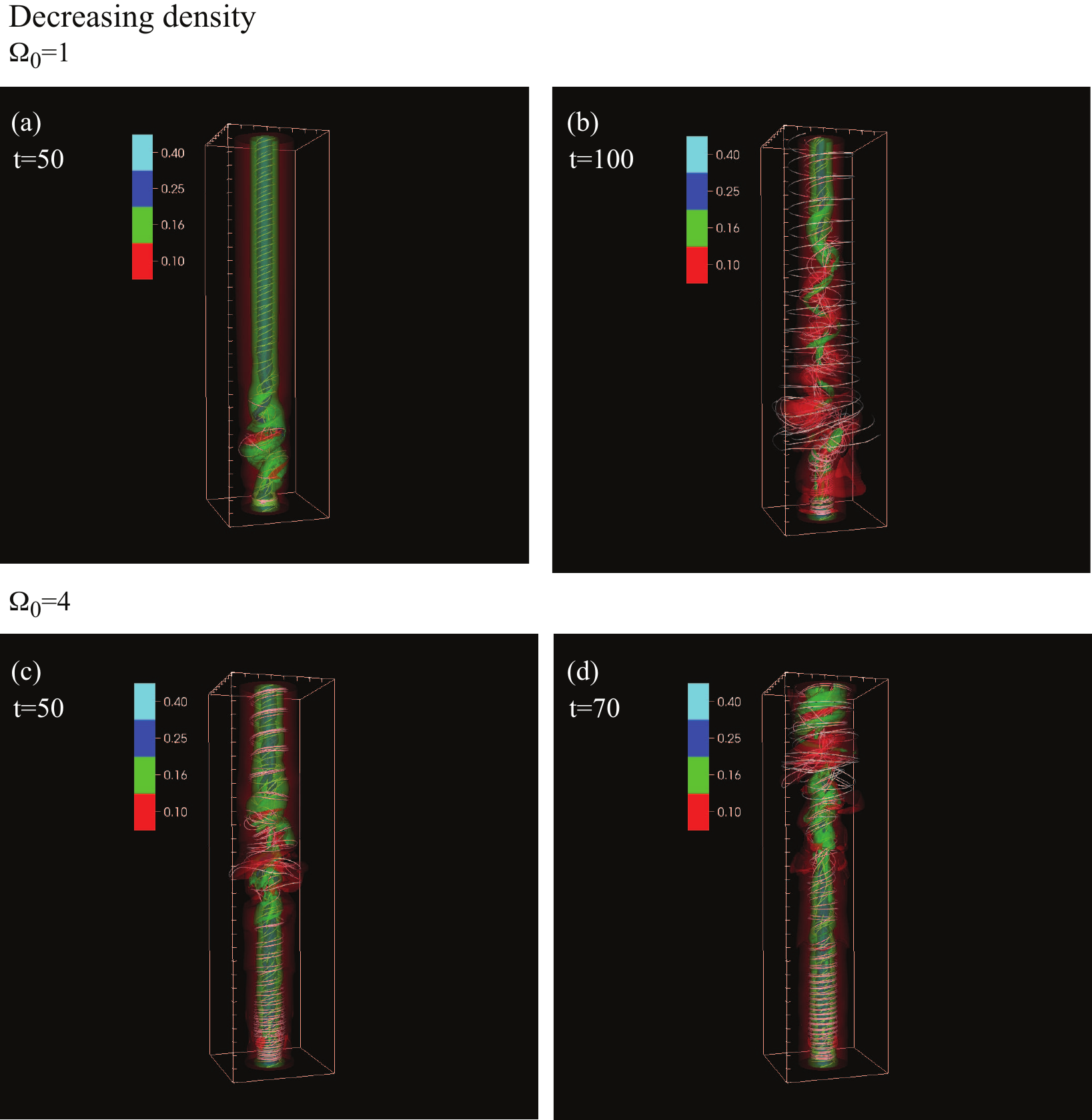}
\end{center}
\caption{Three dimensional density isosurfaces for the decreasing density models. The upper panels {\it (a,b)} show  the $\Omega_{0} = 1$ model
 at $t = 50$ and $100$, while the lower panels {\it (c,d)} show the $\Omega_{0} = 4$ case at $t = 50$
and $70$. The solid lines correspond to the magnetic field lines and the color scales give  the values of the density.}
\label{f2_3Ddec}
\end{figure*}

\begin{figure*}[!ht]
\begin{center}
\includegraphics[width=0.9\textwidth]{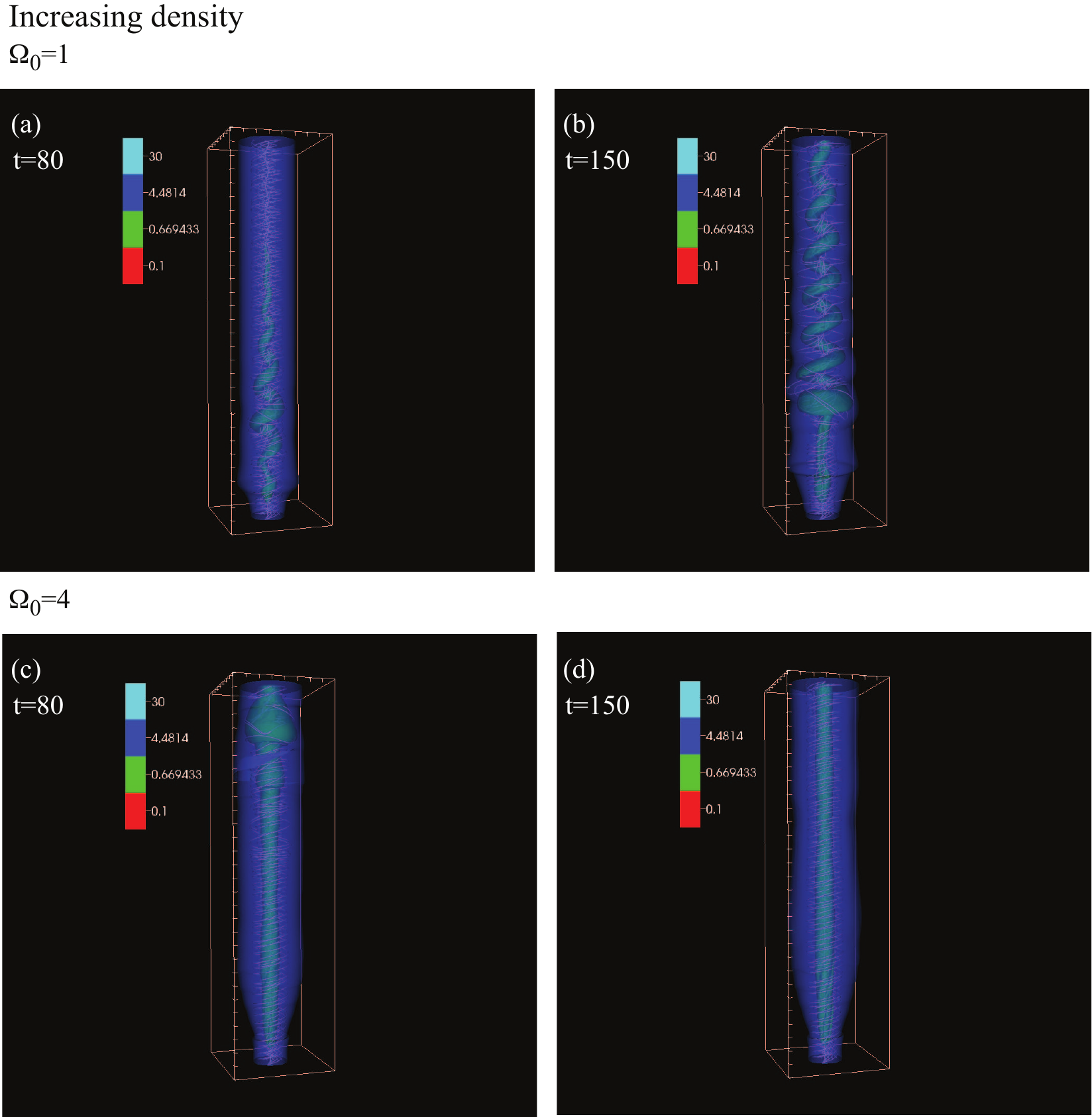}
\end{center}
\caption{Three dimensional density isosurfaces for the increasing density models. The upper panels {\it (a,b)} show the $\Omega_{0} = 1$ case
 at $t = 80$ and $150$, while the lower panels {\it (c,d)} show the $\Omega_{0} = 4$ case at  $t = 80$
 and $150$. Lines and colors are the same as in Figure \ref{f2_3Ddec}.
}
\label{f3_3Dinc}
\end{figure*}

Figure \ref{f2_3Ddec} depicts the $\Omega_{0} = 1$ jet with decreasing density profile at $t = 50$, where
$t$ is in units of $L/c$. It shows that the helical kink develops near the inlet with the largest amplitude at $z \sim 6L$. 
At the later  time $t = 100$, the kink amplitude tends to grow near the jet inlet, but also propagates downstream the jet exciting 
the CD kink instability. The perturbation also grows in time attaining the largest amplitude near the inlet at $z \sim 7-9L = 28-36a$. 
 Also it is found that the precessional perturbation crosses the simulation grid before $t = 100$.
For the $\Omega_{0} = 4$ model with decreasing density in the bottom panels of Figure \ref{f2_3Ddec}, the helical kink also develops 
near the inlet and propagates  downstream  as time increases.
 At  $t = 50$, the results suggest a spatial growth of the CD kink instability with  the largest amplitudes at $z \sim 15-17L = 
60-68a$, well beyond where the largest amplitudes were found for jet with angular velocity $\Omega_{0} = 1$. At the more evolved
time $t = 70$, the kink amplitude increases spatially downstream the jet with the largest amplitudes located far from the inlet at
 $z \sim 20L = 80a$.

For these decreasing density cases, the cylindrical jet with $\Omega_{0} = 1$ exhibits a temporally growing static kink almost 
stationary near the inlet, but  the kink structure propagates downstream nearly with the flow speed. While in the case of 
$\Omega_{0} = 4$, the growing kink structure propagates fast downstream the jet. 

In both cases, the growth rate of the kink instability itself does not  seems to be dependent on the angular velocity amplitude $\Omega_{0}$ as suggested 
by previous periodic box simulations of temporal kink growth (Mizuno et al. 2012), but it does depend on the jet speed. 
It is found that the perturbation propagation speed is greater than the flow speed and the precessional perturbation crosses
 the grid before the final time $t = 100$ and $70$ for $\Omega_{0} = 1$ and $4$, respectively. Organized helical density and magnetic structure
 appear disrupted at the end of the simulation time, at very different distances from the jet inlet. 

Similar to the decreasing density models, the helical kink also develops first near the jet inlet in the increasing density cases, as we can see
in Figure \ref{f3_3Dinc}. In the model with $\Omega_{0} = 1$ at  $t = 150$ the kink amplitude continues to grow near the jet inlet and also 
propagates downstream the jet as in the case with decreasing density profile and same angular velocity.
The largest amplitudes are somewhat farther from the inlet at $z \sim 10L = 40a$ and appear smaller than in the decreasing density case, as we should expect.
For $\Omega_{0} = 4$, the kink amplitude increases spatially and propagates faster downstream the jet, similar to the decreasing density case. 
Here also the precessional perturbation crosses the grid before the final simulated time  $t = 150$. The decreasing density cases exhibit  higher
growth of the kink instability  as reported in spatial growth studies of non-rotating cases by Mizuno et al. (2014). The reason is
higher Alfv\'{e}n speed in the case of decreasing density profile far from the axis (see Figure \ref{f1_1Dinit}). In short, the growth rate of the global 
3D kink structure seems to be nearly independent of the angular velocity amplitude of the jet but propagates faster downstream for larger angular velocity.

\begin{figure*}[!h]
\begin{center}
\includegraphics[width=0.9\textwidth]{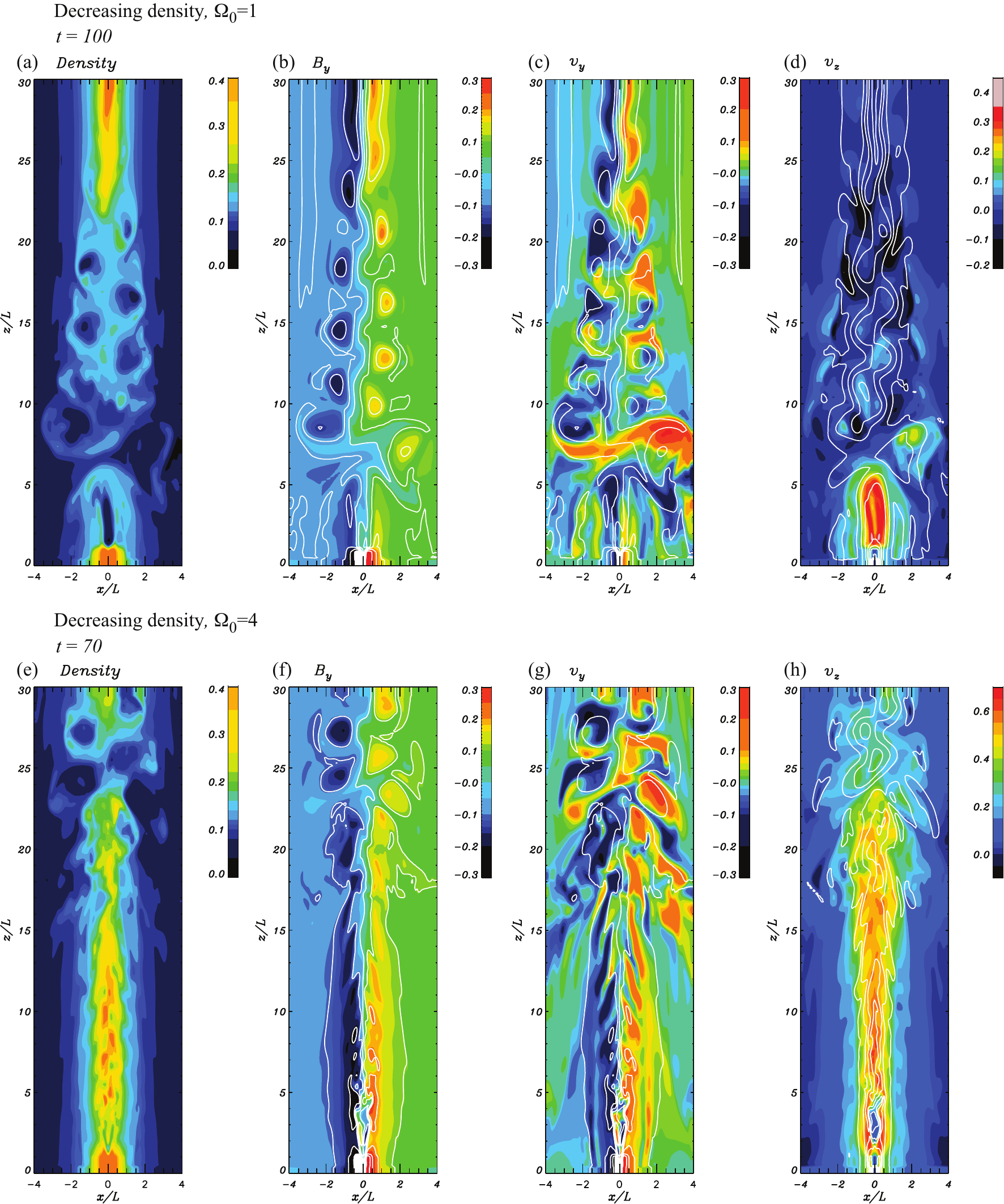}
\end{center}
\caption{Two-dimensional slices for the decreasing density cases at $t = 100$ for $\Omega_{0} = 1$ (upper panels) and $t = 70$ for
$\Omega_{0} = 4$ (lower panels). For both upper and lower panels, the quantities depicted 
from left to right are the density $\rho$ {\it(a,e)}, the azimuthal
magnetic field component $B_{y}$  with  $|B_{y}|$ magnitude contours {\it(b,f)}, the azimuthal $v_{y}$ velocity component with
$|B_{y}|$ magnitude contours {\it(c,g)} and the axial $v_{z}$ velocity component with axial $|B_{z}|$ magnetic field magnitude contours {\it(d,h)}. Slices are in the x-z plane at $y = 0$.
}
\label{f4_2Ddec}
\end{figure*}

\begin{figure*}[!h]
\begin{center}
\includegraphics[width=0.9\textwidth]{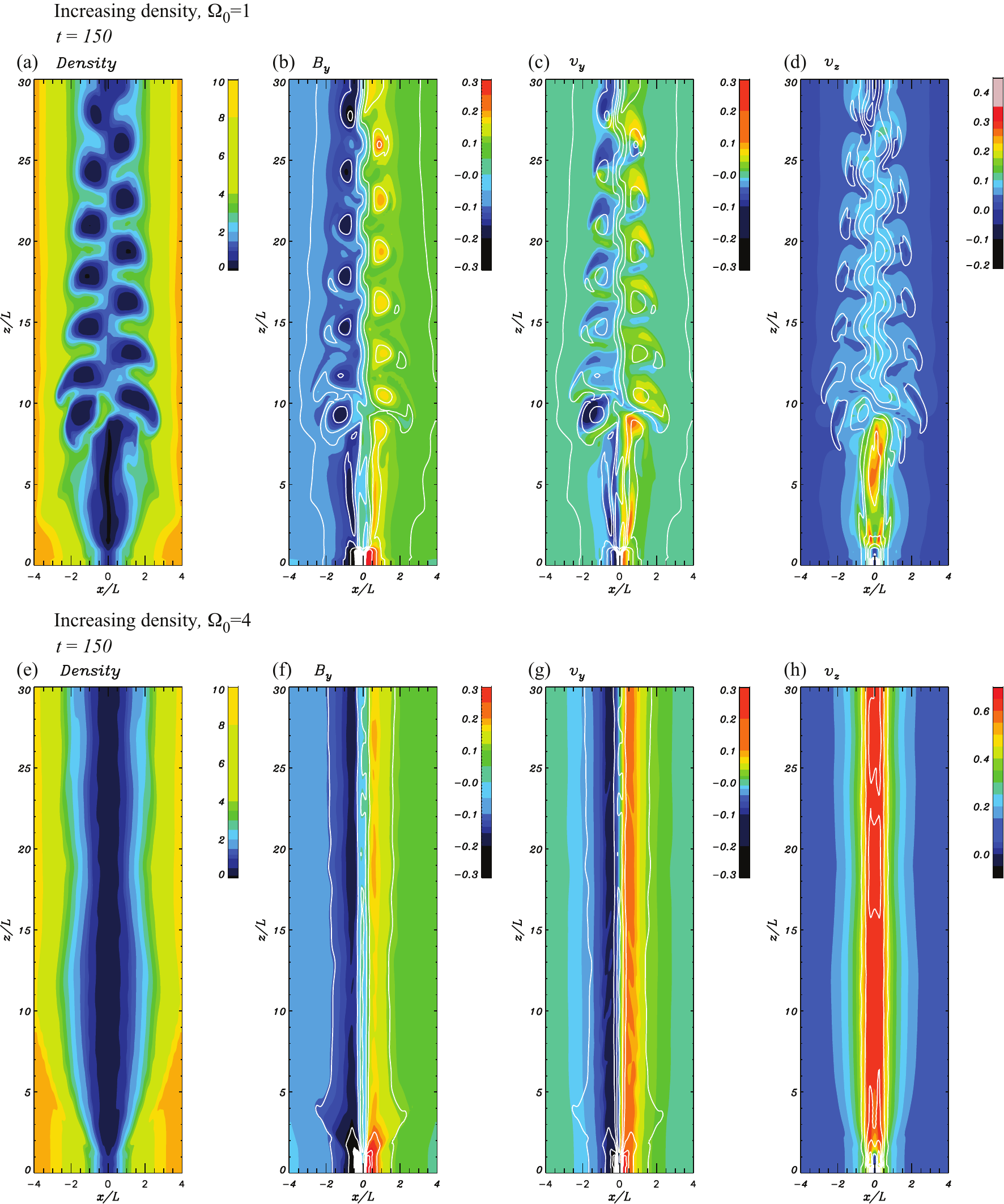}
\end{center}
\caption{Two-dimensional slices for the increasing density cases at $t = 150$ for both $\Omega_{0} = 1$ (upper panels)
and $\Omega_{0} = 4$ (lower panels). For both upper and lower panels it is depicted the same quantities as in Figure \ref{f4_2Ddec}.}
\label{f5_2Dinc}
\end{figure*}

Figure \ref{f4_2Ddec} shows  two-dimensional slices for the decreasing density cases with $\Omega_{0} = 1$ and $4$, for $t = 100$ and $70$, respectively. 
For $\Omega_{0} = 1$, the density and velocity profiles show that the jet flow is strongly distorted as a result of the non-linear growth of the CD kink instability
at $z \sim 5L = 20a$. Within $z < 10L = 40a$, the azimuthal velocity slice shows rapid rotation of the flow as $v_{y} \sim \pm 0.3c$ and 
the axial velocity slice shows acceleration to flow speeds of $v_{z} \sim 0.3c$. Axial velocity of $v_{z} \sim 0.3c$ persists out to 
$z \sim 5L = 20a$. The flow appears primarily as azimuthal  for $z > 5L$  accompanying the helical magnetic field structure that winds around the jet 
plasma column. 

For $\Omega_{0} = 4$ case both the density and axial velocity slices show that after the passage of the kink the jet maintains its initial collimation up to $z \sim 
22L = 88a$ and the axial velocity of $v_{z} \lesssim 0.5c$ persists out to the same location, with some internal structure. The azimuthal velocity
is consistent with a helical flow inside $8a$ and an azimuthal velocity reaching $|v_{y}| \sim 0.2c$. The axial velocity reaches 
up to 0.6c even close to the inlet. The highest speeds are located at $|x/L| \sim L = 4a$. The axial flow appears to 
accompany the helical field beyond $z \sim 23L = 92a$.

Figure \ref{f5_2Dinc} shows similar two-dimensional diagrams for the increasing density models with $\Omega_{0} = 1$ and $4$ at $t = 150$.
For the $\Omega_{0} = 1$ case, both density and axial velocity slices indicate less distortion and less twisted flow out to $z \lesssim 9L = 36a$ 
than the comparable decreasing density cases. Within the same region, the azimuthal velocity slice shows rapid rotation as in the decreasing density case,
with $v_{y} \sim \pm 0.3c$. Similarly, the axial velocity slice indicates acceleration of the flow speed $v_{z} \lesssim 0.3c$ out to $z \sim 9L$,
 but appears less helically twisted than in the decreasing density case. This increasing density model with $\Omega_{0} = 1$ seems to maintain 
a more regular and less distorted structure than the decreasing counterpart.

At the time depicted, the increasing density case with $\Omega_{0} = 4$ (bottom panel of Figure \ref{f5_2Dinc}), the CD kink instability has 
left the computational grid. Both density and axial velocity profiles show a smooth and fast moving jet. The azimuthal magnetic field also appears 
to be quite ordered throughout the computational domain. The azimuthal velocity slice shows rapid rotation with $v_{y} \sim \pm 0.3c$. The axial
velocity indicates acceleration of the flow to speeds up to $v_{z} \sim 0.6c$ close to the axis (at radius $4a$). The flow and magnetic fields 
for this increasing density case are much more ordered than for the decreasing density counterpart.

\begin{figure}[ht!]
\begin{center}
\includegraphics[width=0.5\textwidth]{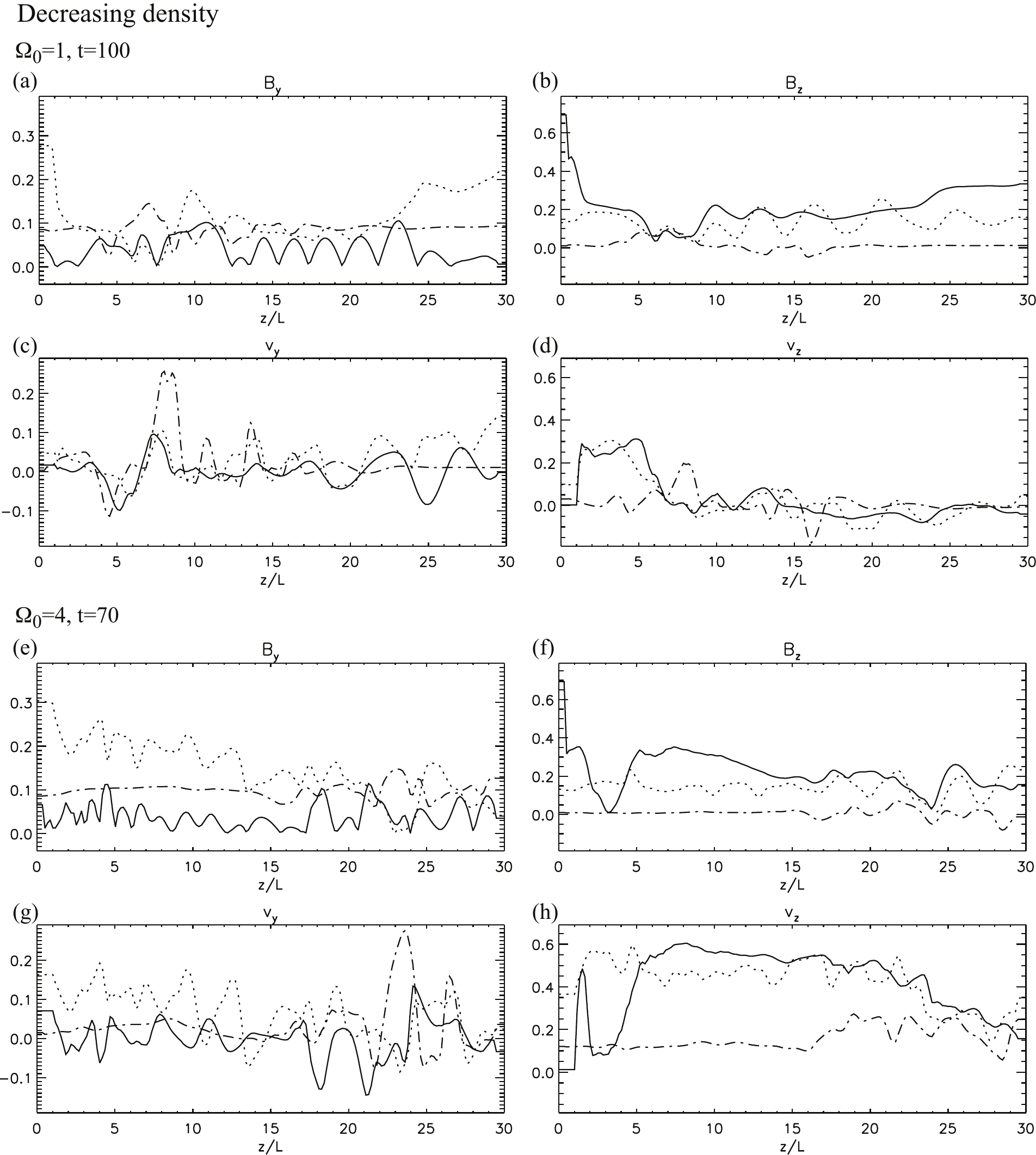}
\end{center}
\caption{One-dimensional cuts parallel to the jet axis for the azimuthal ($B_{y}$, $v_{y}$) and axial ($B_{z}$, $v_{z}$) components of the
magnetic field and velocity  at $y=0$ and $x = 0$ (solid), $L/2 = 2a$ (dotted) and $2L = 8a$ (dash-dotted) for the decreasing density cases
 with $\Omega_{0} = 1$ at $t = 100$ (upper four panels) and $\Omega_{0} = 4$ at $t = 70$ (lower four panels). }
 \label{f6_1Ddec}
\end{figure}

\begin{figure}[h!]
\begin{center}
\includegraphics[width=0.5\textwidth]{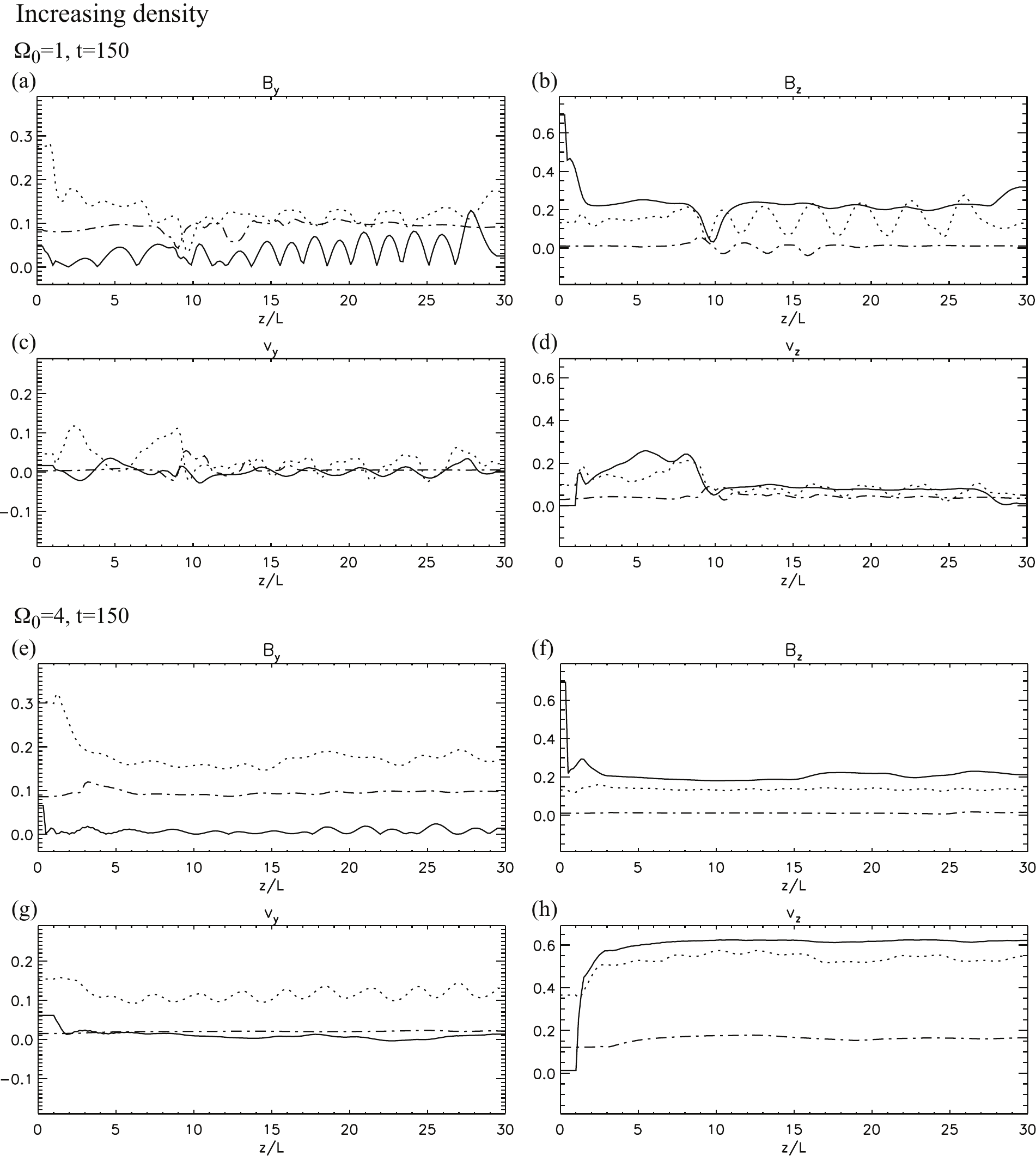}
\end{center}
\caption{One-dimensional cuts parallel the jet axis  for the azimuthal ($B_{y}$, $v_{y}$) and axial ($B_{z}$, $v_{z}$) components of the
magnetic field and velocity  at $y=0$ and $x = 0$ (solid), $L/2 = 2a$ (dotted) and $2L = 8a$ (dash-dotted) for the increasing density
 models with $\Omega_{0} = 1$ at $t = 150$ (upper four panels) and $\Omega_{0} = 4$ at $t = 150$ (lower four panels).}
 \label{f7_1Dinc}
\end{figure}

One can extract further information about the internal structure of the CD kink from  one-dimensional cuts along the $z$-direction.
The panels in Figure \ref{f6_1Ddec} show these cuts for the decreasing density models with $\Omega_{0} = 1$ (D1) and $4$  (D4) made 
at $y=0$ and $x = 0$ (solid), $L/2=2a$ (dotted) and $2L=8a$ (dash-dotted). We see that the azimuthal magnetic field,  $B_{y}$, structure associated 
with the CD kink is more substantial for $x \leq 2a$ for both values of the angular velocity and less significant for $x > 2a$.
Both cases exhibit oscillating patterns at $x = 2a$ with wavelength $\lambda_{k}^{l} \simeq 3.5L$ compatible with the results seen in the 2D slices in Figure \ref{f4_2Ddec}.
The kink structure at shorter wavelength $ \lambda_{k}^{s} = 2L$ is seen in the region $ 14L < z < 24L$ in the case of $\Omega_{0} = 1$
 and $z < 15L$ in the case of $\Omega_{0} = 4$. Internal shorter kink wavelength is seen at $x = 0$. The one dimensional cut for $v_{z}$ indicates 
that the kink grows essentially in one place for $\Omega_{0} = 1$ case, while it propagates downstream for $\Omega_{0} = 4$ case.
The longer wavelength CD kink can be also seen in the oscillations of  the axial magnetic field component, $B_{z}$ for both $\Omega$ cases 
at $x = 2a$. The  axial velocity component, $v_{z}$, also shows oscillating pattern associated with the CD kink at $z \sim 14L$ for $\Omega_{0} = 1$ 
and $z > 30L$ for $\Omega_{0} = 4$.

Figure \ref{f7_1Dinc} shows similar one-dimensional cuts for the increasing density models I1 and I4.
 There is regular oscillating structure in the azimuthal component of the magnetic field, $B_{y}$, associated with the CD kink at $x \leq 2a$ 
for $\Omega_{0} = 1$ (I1), but a smooth structure for $\Omega_{0} = 4$ (I4). As in the decreasing density cases, there is less
 significant or no magnetic structure for  $x > 2a$ for $\Omega_{0} = 1$ and  $\Omega_{0} = 4$, respectively. 
As in the decreasing density case the oscillating structure associated with the CD kink at $x = 2a$  
has wavelength $ \lambda_{k} = 3.5L$ for $\Omega_{0} = 1$. In the case of $\Omega_{0} = 4$, the kink structure has longer wavelength 
$\lambda_{k}^{l} = 3L$ as in the 2D slices shown in Figure \ref{f5_2Dinc}.
Along the axis at $x = 0$, shorter wavelength kink structure  with $\lambda_{k}^{s} = 2L$  is seen in the regions $ 10L < z < 27L$ and  
$z > 30L$ in the case of $\Omega_{0} = 1$ and $4$, respectively. As in the decreasing density models, the kink grows in a single place 
for $\Omega_{0} = 1$, while for $\Omega_{0}= 4$ it propagates downstream and even leaves the computational domain. 
The longer wavelength CD kink can be seen in the oscillations associated with the axial
 magnetic component, $B_{z}$, for $\Omega_{0} = 1$ at $x = 2a$, but not for the $\Omega_{0} = 4$ case.
Indications of the azimuthal motions associated to the CD kink can be identified in the $v_{y}$ velocity component at 
$x = 0$ and $2a$ in the $\Omega_{0} = 1$ case, and at $x = 2a$ in the $\Omega_{0} = 4$ case. 
The axial velocity component ($v_{z}$) also shows oscillatory behaviour associated with the CD kink at $z \sim 10L$ for 
the  $\Omega_{0} = 1$  case. For $\Omega_{0} = 4$, the pattern has left the computational box already.   
This is compatible with the results shown in the three and two dimensional plots (Figs. \ref{f3_3Dinc} \& \ref{f5_2Dinc}).
 
These 1D cuts show maximum axial speeds $v_{z}^{max} \sim 0.3c$, $0.5c$, $0.15c$ and $0.55c$ near the inlet for  D1, D4, I1 and I4 models, respectively. 
For D1 and I1 cases, there is acceleration  mostly along the axis ($x=0$) close to the inlet, as also seen in 2D slices (Figs. \ref{f4_2Ddec} \& \ref{f5_2Dinc}).
At $ x = 8a$,  the azimuthal motion indicates  $v_{y} \sim 0.3c$ near the inlet and far 
away from the inlet for D1 and D4 cases, respectively, while for I1 and I4 cases, it is negligible.
For cases D4 and I4, axial acceleration occurs for $ x \leq 2a$ while the azimuthal motion is seen in 1D cuts at $x = a$ 
and $2a$. Compared to D1 and I1 cases, the axial propagation continues for larger distances with significantly
higher value in D4 and I4 cases, as seen in 2D slices.

According to the KS criterion, the instability develops at $ \lambda_{k} > 2 \pi a $ and a shorter kink wavelength 
$\lambda_{k}^{s} = 6a$ is likely to be seen. Appl et al. (2000) found a fastest growing wavelength $\lambda_{k} \simeq 8.43a$ for 
the static case of constant pitch and uniform density. In all the simulations (D1, D2 and D4 cases) of decreasing density profile, a temporally
 growing kink wavelength of $ \lambda_{k}^{l} \sim 3.5L = 14a$ is identified which is of the order of the wavelength given by wave propagation
 at the jet speed associated with the perturbation frequency. This wavelength is about $67\%$ longer than the predicted fastest growing
 wavelength. A shorter amplitude kink wavelength $\lambda_{k}^{s} \simeq 2L = 8a$ is found to be in the region close to the axis. 
The initial configuration mainly gets distorted because of the nonlinear growth of the longer wavelength. Here, it is seen that the shorter
kink wavelength is of the same order as that of the predicted fastest growing wavelength. The possible reason can be the reconfiguration of 
the magnetic field induced by the precessional perturbation which can reduce the magnetic pitch in the inner part of the jet (see also Mizuno et al. 2012).
The development of an inner shorter wavelength is similar to the periodic grid simulations of rapidly rotating jet and it was 
shown that the inner shorter wavelength appeared where the magnetic pitch was smaller and an outer longer wavelength appeared 
where the pitch parameter was larger (Mizuno et al. 2012). 
In the case of the increasing density profile ( I1, I2 and I4 cases), a spatially growing kink wavelength of similar magnitude 
$ \lambda_{k}^{l} \sim 3.5L = 14a$ to that of the decreasing density is observed. Again, a shorter amplitude kink wavelength 
$ \lambda_{k}^{s} \sim 2L = 8a$ is found to be operating close to the jet axis.

\begin{figure}[!h]
\begin{center}
\includegraphics[width=0.4\textwidth]{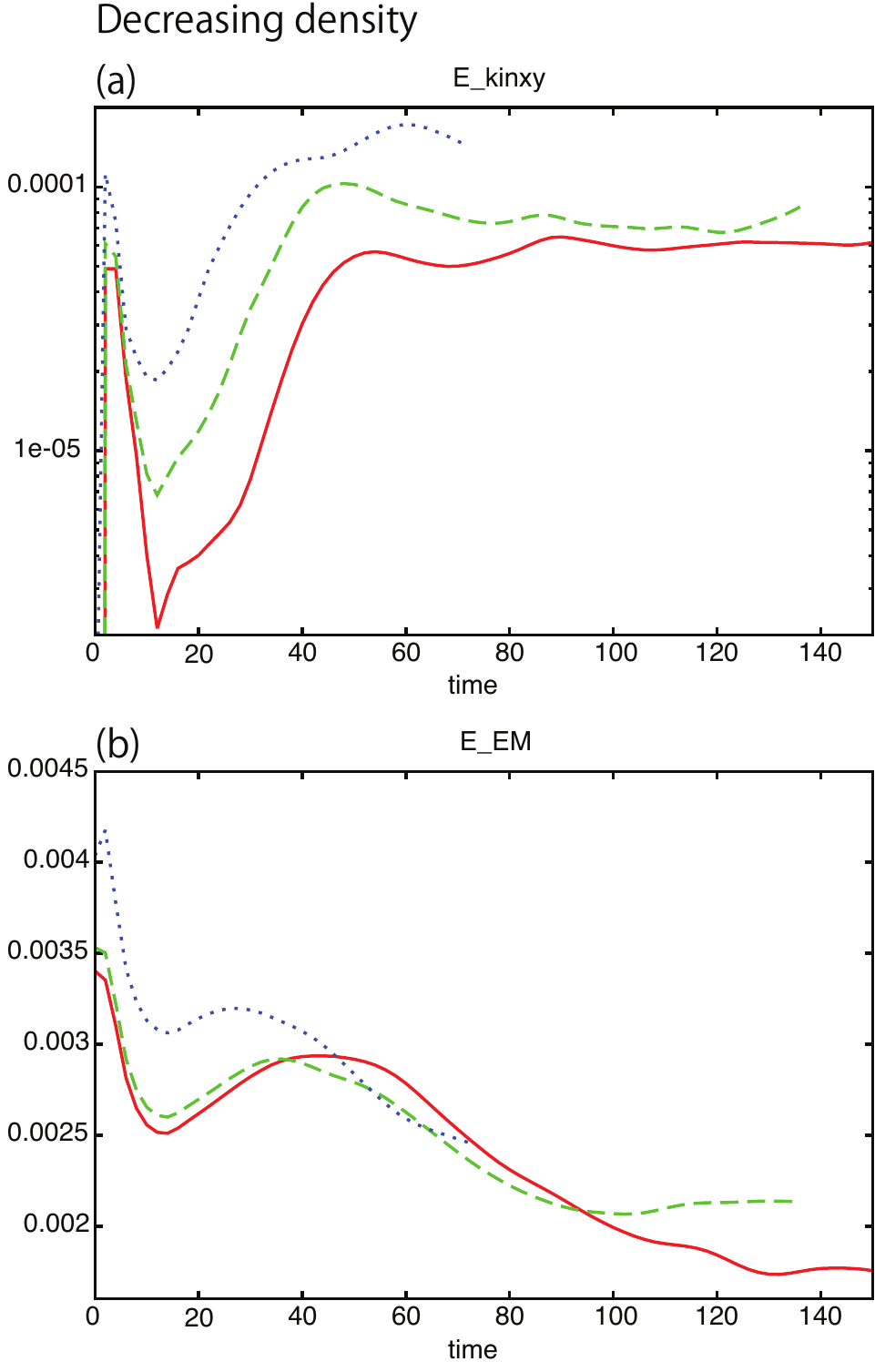}
\end{center}
\caption{Time evolution of volume-averaged $E_{kin,xy}$ (top) and $E_{EM}$ (bottom) for the decreasing density models with
$\Omega_{0} = 1$ (solid red line), $2$ (dashed green line) and $4$ (dotted blue line) within a cylinder of radius $R \le 2L$. $E_{kin,xy}$ is
the integrated kinetic energy transverse to the z-axis and $E_{EM}$ is the integrated total relativistic electromagnetic energy.
}
\label{f8_Eevol_dec}
\end{figure}

\begin{figure}[!h]
\begin{center}
\includegraphics[width=0.4\textwidth]{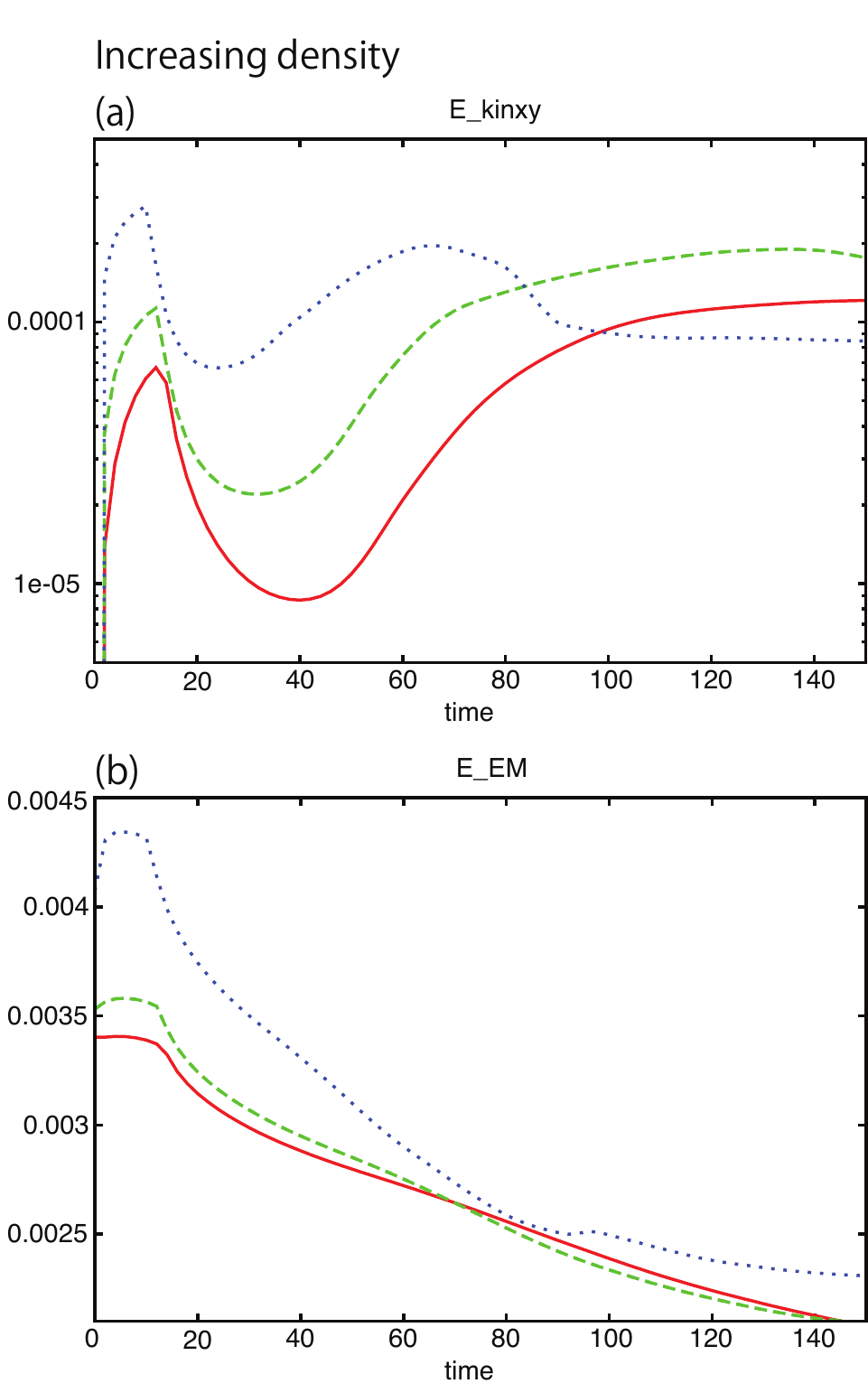}
\end{center}
\caption{Time evolution of volume-averaged $E_{kin,xy}$ (top) and $E_{EM}$ (bottom) for the increasing density models with
$\Omega_{0} = 1$ (solid red line), $2$ (dashed green line) and $4$ (dotted blue line) within a cylinder of radius $R \le 2L$. $E_{kin,xy}$ is
the integrated kinetic energy transverse to the z-axis and $E_{EM}$ is the integrated total relativistic electromagnetic energy.
}
\label{f9_Eevol_inc}
\end{figure}

Figure \ref{f8_Eevol_dec} compares the time evolution of the
 volume-averaged kinetic energy transverse to the z-axis  (see also Mizuno et al. 2012) with the volume-averaged total relativistic electromagnetic energy 
for the decreasing density models with  angular velocities $\Omega_0=1$ (solid red line), $2$ (dashed green line) and $4$ (dotted blue line). 
It is expected that the initial relaxation of the system to cylindrical equilibrium leads to a hump in the kinetic energy plot. The kink 
instability comes into play only after the relaxation finishes. There is an exponential growth from a minimum near $t \sim 10$ to a maximum 
at $t \lesssim 60$. 
Based on the work by Appl et al. (2000), for constant pitch field the estimate of the maximum instability growth rate is given by 
$\Gamma_{max} = 0.133v_{A0}/P_{0}$ where $P_{0}$ is the magnetic pitch at the jet axis (see Figure 1 for values of $P_{0}$ for the different 
cases of initial angular velocity amplitude parameter). Theoretical estimates for the $\Omega_{0}$ = 1, 2 and 4 cases are  found to 
be 0.360$t^{-1}$, 0.384$t^{-1}$ and 0.494$t^{-1}$, respectively. Based on results of Figure \ref{f8_Eevol_dec}, 
the instability growth rates for $\Omega_{0}$ = 1, 2 and 4 are 0.080$t^{-1}$, 0.071$t^{-1}$ and 0.061$t^{-1}$ respectively,
therefore, much smaller than the theoretical estimates of the maximum growth rates.
This is because in the calculation of the volume-averaged energies there are several kink wavelengths that develop along the jet and 
they are possibly camouflaging the maximum predicted growth rates.

The most striking feature in Figure \ref{f8_Eevol_dec} is that the evolution of the relativistic electromagnetic energy is 
anti-correlated with the time evolution of the kinetic energy as the CD kink instability develops at the expense of the magnetic energy.

Similarly, Figure \ref{f9_Eevol_inc}  shows the behaviour of the volume-averaged kinetic and electromagnetic energies for the increasing density cases.
The kinetic energy shows exponential growth from a minimum near $t \sim 40$ to a maximum at $t \lesssim 80$. 
The growth rates are found to be 0.049$t^{-1}$, 0.056$t^{-1}$ and 0.033$t^{-1}$ for $\Omega_{0}$ = 1, 2 and 4, respectively, 
therefore, similar to the results found for the decreasing density models.
Here also the evolution of the electromagnetic energy is opposite to the evolution of the kinetic energy.

We should emphasize that in this work we were able to track the growth of the multiple kink modes whereas in the temporal studies 
of Mizuno et al. (2012), only a single dominant kink or a continuously growing kink structure could be studied.

\begin{figure}[!h]
\begin{center}
\includegraphics[width=0.5\textwidth]{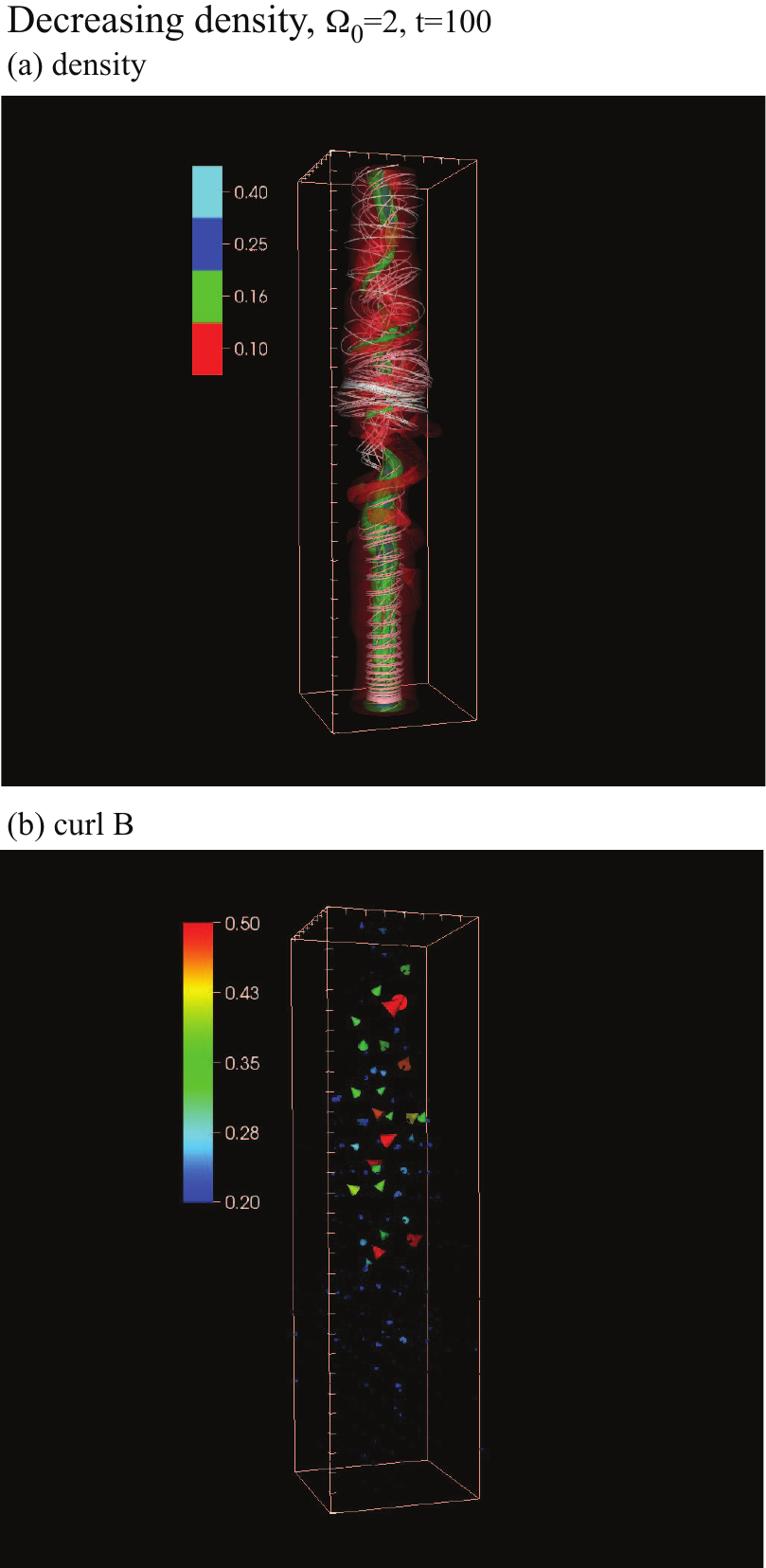}
\end{center}
\caption{Three dimensional density isosurfaces for the decreasing density model with $\Omega_{0} = 2$ (D2) at $t = 100$ (top), and 
the locations of maximum current density, curl $\vec{B}$, which trace the regions where magnetic reconnection may occur, at the same
 time (bottom).
}
 \label{f10_curlB_dec}
\end{figure}

\begin{figure}[!h]
\begin{center}
\includegraphics[width=0.5\textwidth]{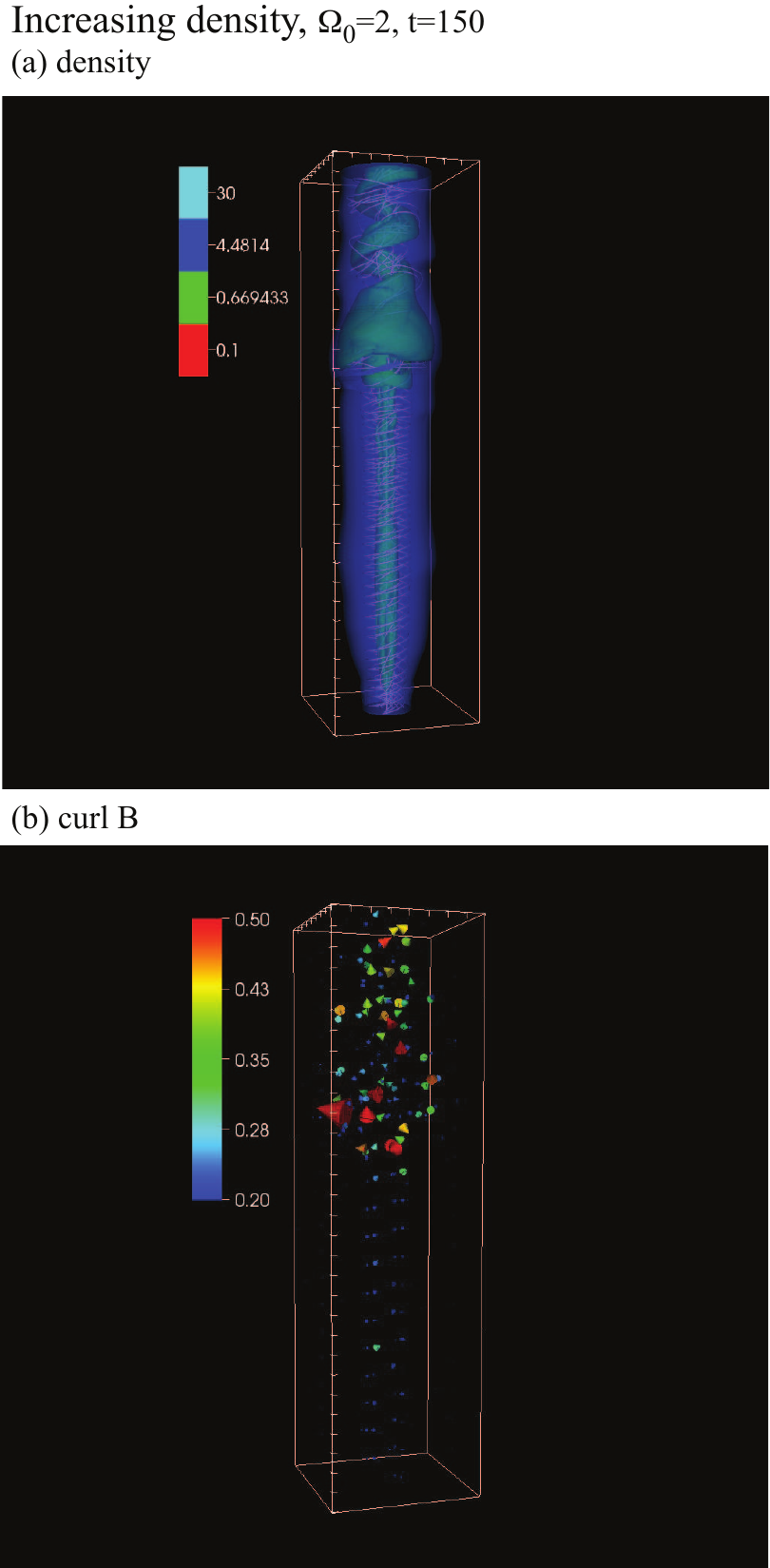}
\end{center}
\caption{The same as in Figure \ref{f10_curlB_dec}, but for the increasing density model with $\Omega_{0} = 2$ (I2) at $t = 150$.
}
\label{f11_curlB_inc}
\end{figure}

\subsection{Identification of magnetic reconnection correlated with the kink instability}

In the previous section we have seen that magnetic energy is being transformed into  kinetic energy as the CD kink instability develops.
In this section, we will try to identify the physical processes behind this behaviour.

Figure \ref{f10_curlB_dec} on top shows the 3D density distribution with  magnetic field lines for the decreasing density model 
 with $\Omega_{0} = 2$ (D2) at  $t = 100$ when  the kink develops downstream the jet with the largest amplitude (located at $z \sim 13L$). 
The  lower panel depicts arrows  that identify the locations where the current density  $\vec{J} \propto \vec{\nabla } \times \vec{B}$ attains the largest intensities. They trace the locations of inversion of the polarity  of the magnetic field lines and therefore, of potential magnetic reconnection sites. 
In fact, magnetic reconnection is expected to occur whenever two magnetic fluxes of opposite polarity approach each other in the presence of finite magnetic 
resistivity. The ubiquitous microscopic Ohmic resistivity is enough to allow for magnetic reconnection, though in this case the rate at which the lines reconnect
 is very slow according to the Sweet-Parker mechanism (e.g., Parker 1979). In the present analysis, we are dealing with relativistic ideal MHD simulations 
with no explicit resistivity. This in principle would prevent us from detecting magnetic reconnection. Nevertheless, in the numerical MHD simulations, 
magnetic reconnection can be excited because of the presence of numerical resistivity. This in turn, could make one to suspect that the identification 
of potential sites of magnetic reconnection in ideal MHD simulations would be essentially a numerical artifact. However, fortunately, based on the
 Lazarian-Vishniac model (see Lazarian \& Vishniac 1999; Kowal et al. 2009; Eyink et al. 2011; Santos-Lima et al. 2010; 2012), 
 the presence of turbulence in real MHD flows is expected to speed up the reconnection to rates nearly as large as the local Alfv\'{e}n speed. This is because of the
 turbulent wandering of the magnetic field lines that allow them to touch each other in several patches simultaneously making reconnection very fast.
 Even in sub-Alfv\'{e}nic flows as in the present case, where the magnetic fields are very strong this {\it fast} reconnection process may be very efficient. 
We see here that the kink instability itself is able to trigger turbulence in the flow as well as the close encounter of magnetic field lines with 
opposite polarity at least in one of the directions, in several regions which are identified by the large increase of the current density in very narrow regions (which are potential sites 
of magnetic discontinuities or current sheets). The numerical resistivity simply mimics the effects of the Ohmic resistivity, but the real and important 
physical effect that may allow for the fast reconnection is the process described above due to the turbulence. This process has been extensively and successfully 
tested numerically in several studies involving ideal MHD simulations of turbulent flows (see e.g., the references above and the reviews of
 Lazarian et al. 2012; Lazarian et al. 2015).

\begin{figure*}[!ht]
\begin{center}
\includegraphics[width=1.0\textwidth]{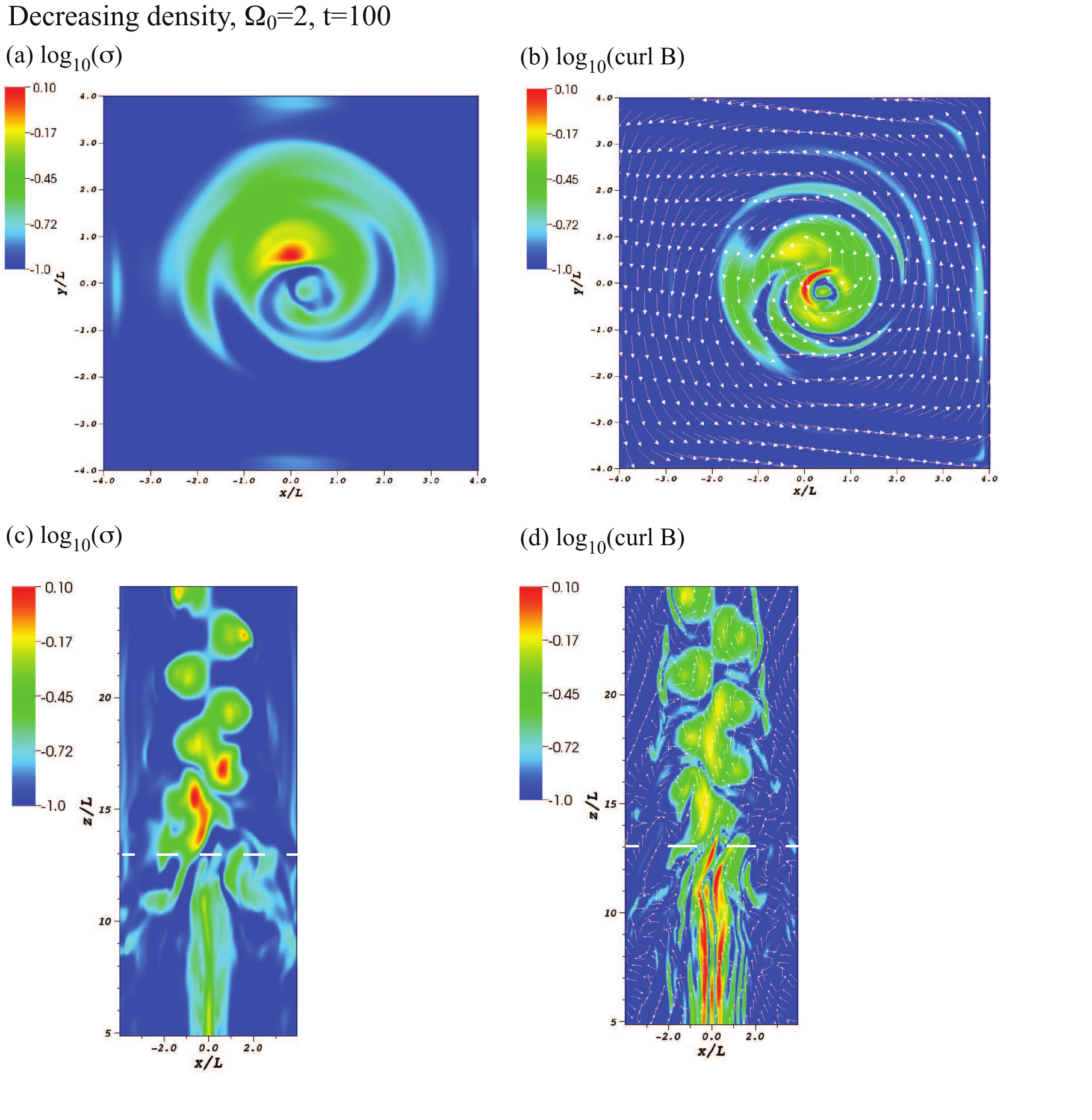}
\end{center}
\caption{Two-dimensional slices for the decreasing density model with $\Omega_{0} = 2$ (D2) at $t = 100$ in the x-y plane at 
$z = 13$ (upper panels) and in the x-z plane at $y = 0$ (lower panels). For both upper and lower panels the quantities depicted 
are the logarithm of the magnetization parameter $\sigma$ where $\sigma = B^{2}/\gamma^2 \rho h$ {\it(a,c)} (left) and the logarithm
 of the current density $curl \vec{B}$ with vectors of the magnetic field superposed to it {\it(b,d)} (right). The white dashed line
 in the x-z diagrams indicate the location along the $z$ axis depicted in the  x-y panels.
}
\label{f12_2DcurlB_dec}
\end{figure*}

\begin{figure*}[!ht]
\begin{center}
\includegraphics[width=1.0\textwidth]{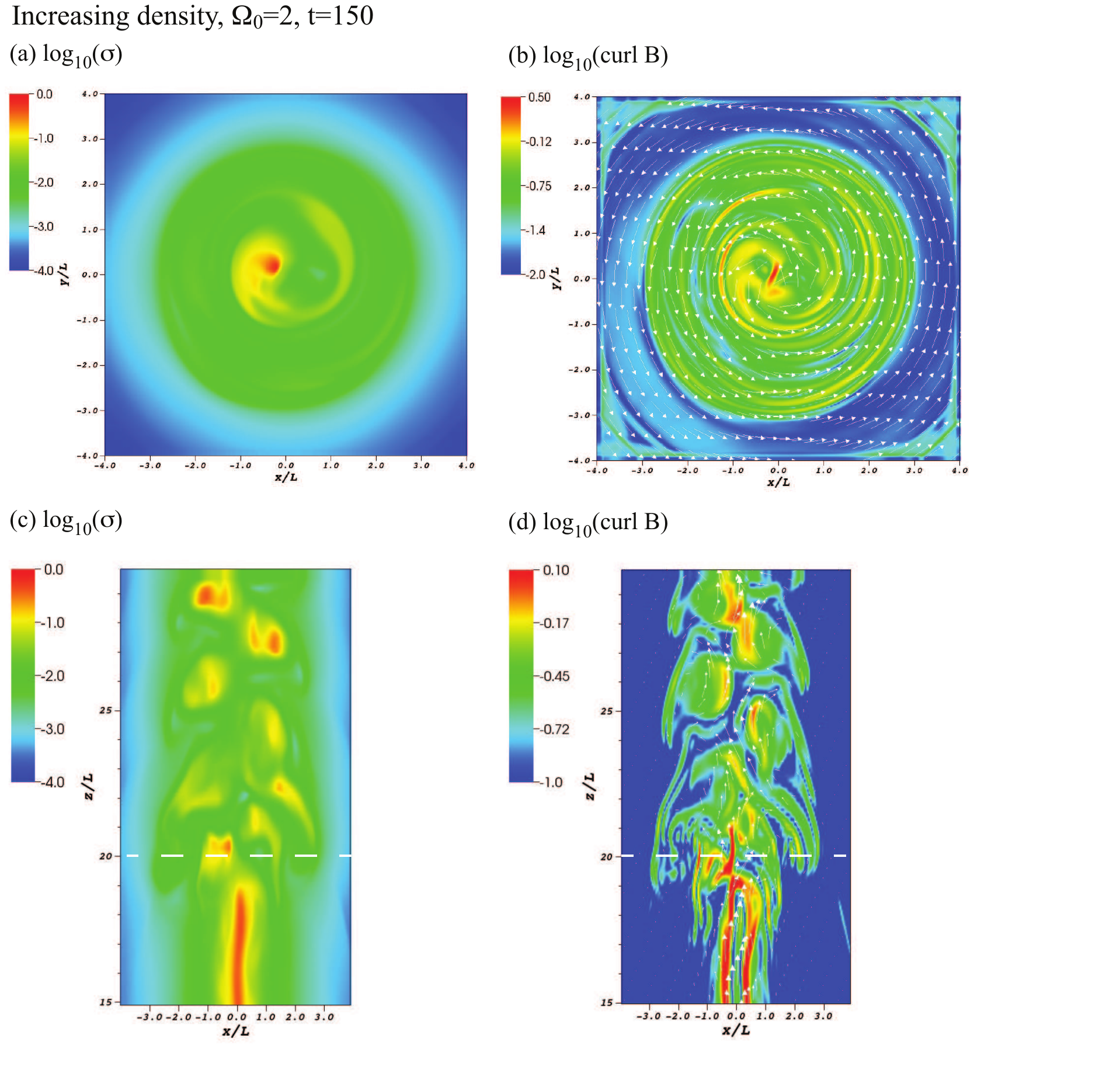}
\end{center}
\caption{Same as in Figure \ref{f12_2DcurlB_dec} but for the model with increasing density with $\Omega_{0} = 2$ (I2) at
 $t = 150$ in x-y plane at $z = 20$ (upper panels) and in x-z plane at $y = 0$ (lower panels).}
\label{f13_2DcurlB_inc}
\end{figure*}


\begin{figure*}[!ht]
\begin{center}
\includegraphics[width=1.0\textwidth]{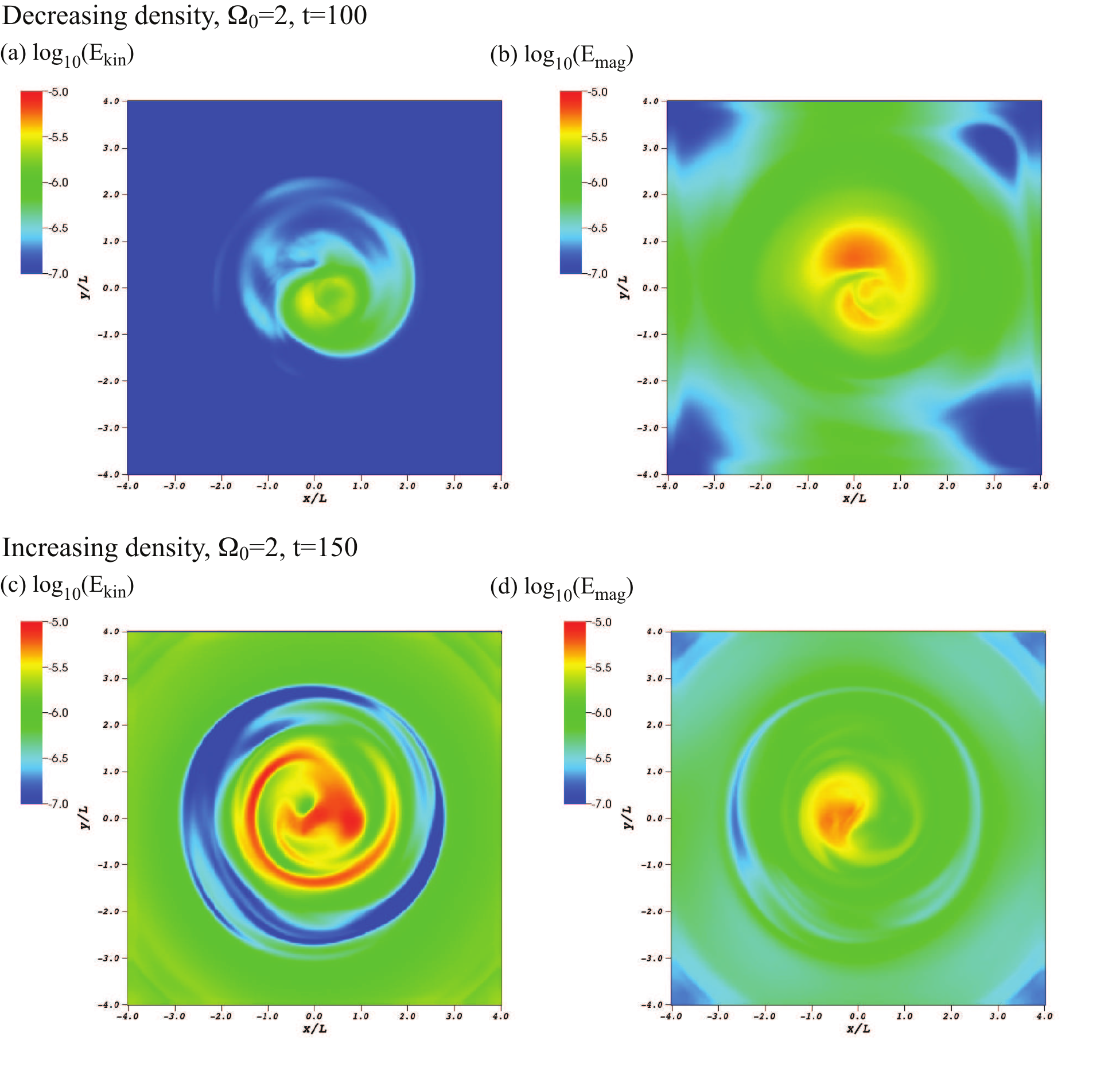}
\end{center}
\caption{ Two-dimensional slices of logarithmic kinetic ({\it left}) and magnetic ({\it right}) energies for the decreasing density model (D2) at $t = 100$ in 
the x-y plane at $z = 13$ as in Figure 12 (upper panels), and for the increasing density model (I2) at $t = 150$ in the xy plane at $z = 20$ as in Figure 13 (lower panels).}
\label{f14_2Dturb}
\end{figure*}


In summary, based on the facts above we may expect a direct correlation between the regions where there is maximum growth of the kink
 instability and regions of enhanced current densities where magnetic reconnection is possibly occurring.
From Figure \ref{f10_curlB_dec}  we clearly see that this is the case, i.e., the location of maximum kink amplitude matches
 the maximum values of $curl \vec{B}$ around $z \geq 13L$. Hence, this  region is also the possible site for maximum energy dissipation
 through magnetic reconnection which can contribute to the increase in kinetic energy of the jet flow and also allow for efficient 
particle acceleration (de Gouveia Dal Pino \& Lazarian 2005; Kowal et al. 2011, 2012; de Gouveia Dal Pino \& Kowal 2015; see also \S 4 below). 

This is also  confirmed by Figure \ref{f12_2DcurlB_dec} where 2D slices of the logarithmic distribution of the current density 
intensity in the x-z plane (at y = 0)  and also in the x-y plane  (at $z = 13L$)  are depicted for the model of Figure 10 (D2).
The figure  also shows  2D slices of the logarithmic distribution of the magnetization parameter  $\sigma = B^{2}/\gamma^2 \rho h$
for the same model.
In the current density x-z  diagram  we identify a complex filamentary structure (contrasting regions of  high and small $|\vec{J}|$ values) 
which  coincides with the regions of the maximum  growth of the kink instability at $z \sim 13L$ and beyond. The distribution of the
 magnetic field vectors is equally more chaotic in this region suffering strong changes in direction between the filaments, therefore
 characterizing regions of magnetic discontinuities (current sheets) and reconnection.  These turbulent regions  contrast  with 
the smoother non-perturbed region below $z \sim 13L$ where the jet is dominated by the helical force-free magnetic field configuration. 
In this region, we see that the current density component ($|\vec{J_z}|$)  is very organized,  as expected for a force-free configuration
 where  $\vec{J_z}$ is simply due to the variation of the regular azimuthal component of the magnetic field 
 $\vec{J_z} \propto  \vec{\nabla} \times \vec{B_\phi}$. In the x-y diagram of current density, we also identify the filamentary structure
 that partially destroys the smooth spiral pattern in this direction. Finally, comparing the current density  with the magnetization
 parameter $\sigma$ diagrams, we  see that regions with  large  intensity values of current density  coincide with  regions where the
 magnetization parameter $\sigma$ is small, therefore strengthening the conclusion that these regions are sites of magnetic energy dissipation. 

 Similarly, Figure \ref{f11_curlB_inc} shows the same trend for the increasing density model  with $\Omega_{0} = 2$ (I2) at the
 terminal simulation time $t = 150$. The kink instability grows above  $z \sim 20L$  which corresponds to the location of the maximum
 values of $curl \vec{B}$, as seen in the lower panel.
In Figure \ref{f13_2DcurlB_inc}, in the 2D current density diagrams corresponding to this model  we can also identify the high intensity
 filaments associated to regions of maximum kink instability and steep variations in the magnetic field direction, characterizing sites 
of  magnetic reconnection,  which are also correlated with regions of small magnetization parameter $\sigma$ (depicted in the same figure). 
These results are also indicative that the regions of kink instability are the sites of reconnection, magnetic energy dissipation and 
jet acceleration. 

 To further stress the conclusions above, Figure \ref{f14_2Dturb} shows  2D slices of the logarithmic distribution of the kinetic 
(left panels) and magnetic  (right panels) energies in the xy plane at the same positions depicted in Figures \ref{f12_2DcurlB_dec} and 
\ref{f13_2DcurlB_inc} for the heavy (D2) and light (I2) jets, respectively. In both cases we see that the regions near the center that have
 been identified as reconnection sites  have larger kinetic energy than the
 surroundings, while the magnetic energy is smaller in these regions, indicating the conversion of the magnetic into kinetic energy.
Fast magnetic reconnection induced by turbulence has been previously tested numerically by Kowal et al. (2009) considering large scales 3D current 
sheets with injected steady state, incompressible turbulence. These authors measured the reconnection rate  directly from their simulations considering
 the effective electric field in the plasma ($\vec{v} \times \vec{B}$; see their eq. 13).    
Recently Takamoto et al. (2015) extended this numerical work to the relativistic regime considering both incompressible and compressible flows and found that 
the reconnection rate is  independent of the plasma resistivity as in the  non-relativistic case. They measured reconnection rate values between $(0.01 - 0.1)V_A$
depending on the  magnetization parameter $\sigma$ of the flow. For incompressible plasmas with  $\sigma=0.1$ and $1$, as in the present work, they found values 
$\sim(0.05-0.06)V_A$ and $\sim(0.03-0.05)V_A$ respectively. Following Kowal et al. (2009), we also used their eq. (13) to make a quite rough estimate of the 
reconnection rate in the magnetic reconnection sites identified in the heavy and light jets in Figures \ref{f12_2DcurlB_dec} to \ref{f13_2DcurlB_inc}  above.
Considering the values of the components of $\vec{B}$ and $\vec{v}$ in these regions, we obtained reconnection rates $\sim 0.05 V_A$ for both jets, which are 
comparable to the values obtained by Takamoto et al. (2015). This further indicates that the turbulence due to the CD kink instability in the relativistic jets 
is able to induce fast magnetic reconnection.

\section{Discussion and concluding remarks}

In this work we studied the influence of  jet rotation  on the spatial development of the CD kink instability using a non-periodic 
computational box. In our  spatial-growth simulations and also in previous temporal-growth simulations of the CD kink (e.g., Mizuno et al. 2012),
 the setup is built for the analysis of the structure of Poynting-flux dominated jets (Lyubarsky 2009). In the comoving frame, the poloidal
 and toroidal fields are comparable for cylindrical equilibrium configurations. When the Alfv\'{e}n crossing time is smaller than the
 proper propagation time (if the jet is narrow enough), the jet relaxes to a local equilibrium configuration. Such jets can be subject
 to the kink instability because the characteristic instability timescale is larger than the Alfv\'{e}n crossing time. It is to be
 noted that in relativistic jet flows, when the toroidal field exceeds the poloidal one, the hoop stress is counter-balanced by the
 electric force and the poloidal magnetic field pressure gradient where the centrifugal force and the gas pressure remains small.
 This is the reason why well beyond the Alfv\'{e}n point, the flow can still remain Poynting-flux dominated and 
the flow structure is close to a cylindrical configuration. 

The previous study of temporal growth of the CD kink instability used a periodic computational box and uniform spatial 
perturbation of the jet across the computational box domain. For a radially decreasing density profile,  Mizuno et al. (2012)
 found that for an angular velocity  $\Omega_{0} = 1$ and magnetic helicity $\alpha = 1$, the displacement of the 
initial force-free helical magnetic field resulting from the growth of the CD kink instability leads to a helically twisted 
magnetic filament  around density isosurfaces near the axis associated with the $n = 1$ kink mode wavelength. 
In the nonlinear phase, the helically distorted density structure showed continuous transverse growth and propagated along the flow direction.
In the case of $\Omega_{0} = 2$ and $4$ models, development of both the $n = 1$ and $2$ kink mode wavelengths were observed.  

Here in all models corresponding to $\Omega_{0} = 1$, $2$ and $4$, different kink wavelengths are seen at different positions along the transverse 
x-direction. It is seen that the magnetic pitch strongly decreases in the innermost region of the jet as the angular velocity amplitude
increases. Moreover, it is detected the growth of different axial kink wavelengths since this depends on the magnetic pitch.
Eventually the kink wavelength with maximum growth becomes dominant at each radial position.

However the magnetic pitch remains essentially the same outside the jet core in all cases because of the initial constant magnetic pitch in a force-free 
magnetic field. Thus similar long kink wavelengths are seen outside of the jet core. Besides, short kink wavelengths usually do not evolve. 
On the other hand, at a particular radial location, mostly long axial kink wavelengths can grow which have lower growth rates than shorter kink wavelengths 
inside the jet core. As a result the longer kink wavelengths outside the jet core interact with the shorter kink wavelengths inside the jet core
at later simulation times and further downstream the axial region when the shorter kink wavelengths have already entered the non-linear phase and this would 
accelerate the jet disruption.
The evolution of the volume-averaged kinetic energy transverse to the z-axis shows an exponential growth, in agreement with the general
theoretical estimate of the maximum instability growth rate. The evolution of the total volume-averaged relativistic electromagnetic
energy is opposite to the evolution of kinetic energy, similar to results obtained in Mizuno et al. (2012).

In the present work, the CD kink instability seems to be stabilized in the case of a radially increasing density profile compared to the decreasing 
density counterpart. The same effect was also reported in Mizuno et al. (2014), but considering a spatial-growth study of non-rotating jets. 
An inner high-speed, low-density jet spine surrounded by a denser and slower sheath, as suggested by observations 
(e.g., Boccardi et al. 2015 and references therein) and GRMHD simulations of relativistic jet formation (e.g., 
McKinney 2006) is compatible with the scenario above of the increasing density models. Thus we may expect the development of 
a CD kink in the jet, but it may saturate as a result of a radially increasing density structure as shown, e.g.,  in 
McKinney \& Blandford (2009) and this may also explain the non-destructive kink structures often observed in relativistic jets. 

Our results also suggest that the Poynting-flux  dominated jets are accelerated up to approximate equipartition
 between the kinetic and magnetic energies and then the regular flow may be disrupted by the kink instability. Such a
scenario is valid for jets satisfying the causality condition which applies very well to the case of AGN jets (e.g., Pushkarev et al. 2009).
Though the causality condition seems to be violated in GRB jets, the causally connected regions inside GRB jets can undergo local 
internal kink instability as shown in recent work by Bromberg \& Tchekhovskoy (2015) where the instability can cause small-angle magnetic reconnection (see also Tchekhovskoy \& Bromberg 2015).

Shocks and magnetic reconnection have been considered as possible candidates for powering the jet emissions in AGNs and GRBs.
Recent fully-kinetic particle-in-cell simulations have shown that shock models are unlikely to account for the jet emission 
(Sironi et al. 2015). Shocks are not able to accelerate particle far beyond the thermal energy, however, magnetic reconnection can 
deposit more than 50 per cent of the dissipated energy into non-thermal leptons as long as the energy density of the magnetic field
in the bulk flow is larger than the rest mass-energy density. Several GRB prompt emission models have been proposed based on the role 
of magnetic reconnection in the GRB jets, like the  magnetic heating photosphere model (Giannios 2008), the  internal collision-induced 
 magnetic reconnection  turbulent model   (ICMART; Zhang \& Yan 2011) based on the Lazarian-Vishniac model of fast reconnection
triggered by turbulence and the particle acceleration Fermi mechanism by reconnection (de Gouveia Dal Pino \& Lazarian 2005);
and the magnetic reconnection switch model (McKinney \& Uzdensky 2012).

In the study presented here, we imposed a precessional perturbation at the inlet of a rotating cylindrical jet that leads the displacement of the 
helical magnetic field configuration and induces the CD kink instability which in turn leads to magnetic reconnection and energy dissipation, 
as suggested by the results of Figures \ref{f8_Eevol_dec} to \ref{f13_2DcurlB_inc}. This goes to the kinetic energy of the flow and may also 
accelerate particles through a first-order Fermi process in the magnetic reconnection regions (e.g. de Gouveia Dal Pino \& Lazarian 2005; Kowal et al. 2011; 2012). 
A recent study injecting test particles in an MHD relativistic jet has evidenced that the acceleration by magnetic reconnection may be dominating in the more 
turbulent magnetically dominated regions of the flow, like the cocoon that envelopes the jet, and is a competing process with the acceleration at the internal
 shocks along the beam and at the jet head (de Gouveia Dal Pino \& Kowal 2015).  Further studies considering test particles in relativistic MHD jet 
simulations are required in order to understand better the energy dissipation, the relevant acceleration mechanisms and the non-thermal emission processes
 in such systems. The inclusion of the particle feedback on the plasma and vice-versa may be also implemented in such studies considering an MHD-PIC approach 
(e.g., Bai et al. 2015).

\acknowledgments
C.B.S. acknowledges support from  the Brazilian agency FAPESP ($2013/09065-8$) and  E.M.G.D.P. 
 from FAPESP (2013/10559-5)  and CNPq (306598/2009-4) grants. 
 Y.M. acknowledges support from the ERC Synergy Grant ``BlackHoleCam - Imaging the Event Horizon of Black Holes'' (Grant 610058).
 This work has made use of the computing facilities of the Laboratory of Astroinformatics (IAG/USP, NAT/Unicsul),
 whose purchase was made possible by FAPESP (grant 2009/54006-4). The authors are indebted to Prof. Luciano Rezzolla  for his useful 
comments on this work.
C.B.S. would like to thank Luis H.S. Kadowaki at IAG-USP for discussions regarding use of VisIt tool for Figures in this work.
 VisIt is a publicly available visualization and graphical analysis tool for data defined on 2D and 3D meshes made available by 
Lawrence Livermore National Laboratory, USA.


\begin{thebibliography}{}

\bibitem[Abramowicz \& Fragile (2013)]{abr13} Abramowicz, M. A., \& Fragile, P. C. 2013, LRR, 16, 1

\bibitem[Aloy \& Rezzolla(2003)]{alo03} Aloy, M. A., \& Rezzolla, L. 2006, \apjl, 640, L115

\bibitem[Anjiri et al. (2014)]{anj14} Anjiri, M., Mignone, A., Bodo, G., \&  Rossi, P. 2014, \mnras, 442, 2228

\bibitem[Appl et al. (2000)]{app00} Appl, S., Lery, T., \& Baty, H. 2000, \aap, 355, 818

\bibitem[Bai et al.(2015)]{bai15} Bai, X.-N., Caprioli, D., Sironi, L., \& Spitkovsky, A.\ 2015, \apj, 809, 55 

\bibitem[Bateman (1978)]{bat78} Bateman, G. 1978, MHD Instabilites (Cambridge, MA: MIT Press)

\bibitem[Baty (2005)]{bat05} Baty, H. 2005, \aap, 430, 9

\bibitem[Baty \& Keppens (2002)]{bat02} Baty, H. \& Keppens, R. 2002, \aap, 580, 800

\bibitem[Beckwith \& Stone (2011)]{bec11} Beckwith, K., \& Stone, J. M. 2011, \apjs, 193, 6

\bibitem[Begelman (1998)]{beg98} Begelman, M. C. 1998, \apj, 493, 291

\bibitem[Begelman et al. (1980)]{beg80} Begelman, M. C., Blandford, R. D., \& Rees, M. J. 1980, \nat, 287, 307

\bibitem[Begelman et al. (1984)]{beg84} Begelman, M. C.,  Blandford, R. D., \& Rees, M. J. 1984, RvMP, 56, 255 

\bibitem[Beskin (2010)]{bes10} Beskin, V. S. 2010, PhyU, 53, 1199  

\bibitem[Bisnovatyi-Kogan \& Lovelace (2001)]{bis01} Bisnovatyi-Kogan, G.S., \& Lovelace, R.V.E. 2001, NewAR, 45, 663 

\bibitem[Blandford \& Payne (1982)]{bla82} Blandford, R. D., \& Payne, D. G. 1982, \mnras, 199, 883

\bibitem[Blandford \& Znajek (1977)]{bla77} Blandford, R. D., \& Znajek, R. L. 1977, \mnras, 179, 433

\bibitem[Bromberg \& Tchekhovskoy (2015)]{bro15} Bromberg, O., \& Tchekhovskoy, A. 2015, arXiv:1508.02721

\bibitem[Boccardi et al. (2015)]{boc15} Boccardi, B., Krichbaum, T.P., Bach, U., \& Mertens, F. 2015, \aap, in press (arXiv:1509.06250)

\bibitem[de Gouveia Dal Pino(2005)]{dgdp05} de Gouveia Dal Pino, E.~M. 2005, Advances in Space Research, 35, 908

\bibitem[de Gouveia Dal Pino \& Kowal(2015)]{dgdp_etal_15} de Gouveia Dal Pino, E.~M., \& Kowal, G. 2015, in Magnetic Fields in Diffuse Media, Astrophysics and Space Science Library, A. Lazarian. E. de Gouveia Dal Pino, C. Melioli  (eds.),  407, 373

\bibitem[de Gouveia Dal Pino \& Lazarian(2005)]{dgdp_lazarian_05} de Gouveia Dal Pino, E.M., \& Lazarian, A. 2005, \aap, 441, 845

\bibitem[Del Zanna et al. (2007)]{del07} Del Zanna, L., Zanotti, O., Bucciantini, N., \& Londrillo, P. 2007, \aap, 473, 11

\bibitem[Eyink et al. (2011)]{2011ApJ...743...51E} Eyink, G.~L., Lazarian, A., \& Vishniac, E.~T.\ 2011, \apj, 743, 51 

\bibitem[Gammie et al. (2003)]{gam03} Gammie, C. F., McKinney, J. C. \& Toth, G. 2003, \apj, 589, 444

\bibitem[Giannios (2008)]{gia08} Giannios, D. 2008, \aap, 480, 305

\bibitem[Giannios(2010)]{2010MNRAS.408L..46G} Giannios, D.\ 2010, \mnras, 408, L46 

\bibitem[Giannios \& Spruit (2006)]{gia06} Giannios, D., \& Spruit, H. C. 2006, \aap,  450, 887 

\bibitem[Girolleti et al. (2004)]{gir04} Giroletti, M., Giovannini, G., Feretti, L., et al. 2004, \apj, 600, 127
	
\bibitem[Gomez et al. (2001)]{gom01} Gomez, J. -L., Marscher, A. P., Alberdi, A., Jorstad, S. G., \& Agudo, I. 2001, \apj, 561, L161 

\bibitem[Granot et al. (2015)]{gra15} Granot, J., Piran, T., Bromberg, O., Racusin, J. L., \& Daigne, F. 2015, SSRv, 191, 471

\bibitem[Hardee (2013)]{har03} Hardee, P. E. 2013, in EPJ Web of Conferences, The Innermost Regions of Relativistic Jets
                                and Their Magnetic Fields, ed. J. L. Gomez, 61, 02001

\bibitem[Hsu \& Bellan (2002)]{hsu02} Hsu, S. C., \& Bellan, P. M. 2002, \mnras, 334, 257

\bibitem[Hsu \& Bellan (2005)]{hsu05}Hsu, S. C., \& Bellan, P. M. 2005, PhPl, 12, 2103

\bibitem[Istomin \& Pariev (1994)]{iso94} Istomin, Y. N., \& Pariev, V. I. 1994, \mnras, 267, 629

\bibitem[Istomin \& Pariev (1996)]{iso96} Istomin, Y. N., \& Pariev, V. I. 1996, \mnras, 281, 1

\bibitem[Kato et al. (2008)]{kat08} Kato, S., Fukue, J. \& Mineshige, S. 2008, Black-Hole Accretion Disks : Towards a
                                    New Paradigm (2nd ed., Kyoto: Kyoto Univ. Press)

\bibitem[Keppens et al. (2012)]{kep12} Keppens, R., Meliani, Z., van Marle, A. J., et al. 2012, JCoPh, 231, 718 

\bibitem[Komissarov (1997)]{kom97} Komissarov, S. S. 1997, Phys. Lett. A, 232, 435

\bibitem[Kovalev et al. (2007)]{kov07} Kovalev, Y. Y., Lister, M. L., Homan, D. C., \& Kellermann, K. I. 2007, \apj, 668, L27

\bibitem[Kadowaki et al.(2015)]{2015ApJ...802..113K} Kadowaki, L.~H.~S., de Gouveia Dal Pino, E.~M., \& Singh, C.~B.\ 2015, \apj, 802, 113 

\bibitem[Kowal et al.(2009)]{Kowal_etal_09} Kowal, G., Lazarian, A., Vishniac, E.~T., Otmianowska-Mazur, K., 2009, \apj, 700, 63

\bibitem[Kowal et al.(2011)]{Kowal_etal_11} Kowal, G., de Gouveia Dal Pino, E.M., \& Lazarian, A. 2011, \apj, 735, 102

\bibitem[Kowal et al.(2012)]{Kowal_etal_12} Kowal, G., de Gouveia Dal Pino, E.M., \& Lazarian, A. 2012, \prl, 108, 241102

\bibitem[Kumar \& Zhang (2015)]{kum15} Kumar, P., \& Zhang, B. 2015, PhR, 561, 1 

\bibitem[Lazarian \& Vishiniac(1999)]{LV_99} Lazarian, A., \& Vishniac, E., 1999, \apj, 517, 700

\bibitem[Lazarian et al.(2012)]{2012SSRv..173..557L} Lazarian, A., Vlahos, L., Kowal, G., Yan, H.; Beresnyak, A. \& de Gouveia Dal Pino, E. M. \ 2012, \ssr, 173, 557 

\bibitem[Lazarian et al.(2015)]{Lazetal15} Lazarian, A.,  Kowal, G.,   Takamoto, M.,  de Gouveia Dal Pino, E. M. \& J. Cho 2016, Magnetic Reconnection, W. Gonzalez, E. Parker (eds.), Astrophysics and Space Science Library 427, DOI 10.1007/978-3-319-26432-5\_11

\bibitem[Laurent et al. (2011)]{lau11} Laurent, P., Rodriguez, J., Wilms, J., et al. 2011, Sci, 332, 438

\bibitem[Lery \& Baty (2000)]{ler00} Lery, T., Baty, H., \& Appl, S. 2000, \aap, 355, 1201

\bibitem[Li (2000)]{lil00} Li, L.-X. 2000, \apj, 531, L111

\bibitem[Lobanov \& Zensus (2001)]{lob01} Lobanov, A. P., \& Zensus, J. A. 2001, Sci, 294, 128

\bibitem[Lobanov et al. (2003)]{lob013} Lobanov, A., Hardee, P., \& Eilek, J. 2003, NewSciRv, 47, 629  

\bibitem[Lyubarskii (1999)]{lyu99} Lyubarskii, Y. E. 1999, \mnras, 308, 1006

\bibitem[Lyubarky (2009)]{lyu09} Lyubarsky, Y. 2009, \apj, 698, 157

\bibitem[Marti-Vidal et al. (2015)]{mar15} Marti-Vidal, I., Muller, S., Vlemmings, W., Horellou, C., \& Aalto, S. 2015, Sci, 348, 311

\bibitem[McKinney (2006)]{mck06} McKinney, J. C. 2006, \mnras, 368, 1561

\bibitem[McKinney \& Blandford (2009)]{mck09} McKinney, J. C., \& Blandford, R. D. 2009, \mnras, 394, L126

\bibitem[McKinney \& Uzdensky (2012)]{mck12} McKinney, J.C., \& Uzdensky, D. A. 2012, \mnras, 419, 573

\bibitem[Mignone et al. (2007)]{mig07} Mignone, A., Bodo, G., Massaglia, S., et al., 2007, \apjs, 170, 288

\bibitem[Mignone et al. (2013)]{mig13} Mignone, A., Striani, E., Tavani, M., \& Ferrari, A., 2013, \mnras, 436, 1102

\bibitem[Mignone et al. (2010)]{mig10} Mignone, A., Tzeferacos, P., \& Bodo, G. 2010, JCoPh, 229, 5879

\bibitem[Mizuno et al.(2011)]{miz11} Mizuno, Y., Hardee, P. E., \& Nishikawa, K.-I. 2011, \apj, 734, 19 

\bibitem[Mizuno et al.(2014)]{miz14} Mizuno, Y., Hardee, P. E., \& Nishikawa, K.-I. 2014, \apj, 784, 167

\bibitem[Mizuno et al.(2009)]{miz09} Mizuno, Y., Lyubarsky, Y., Nishikawa, K.-I., \& Hardee, P. E. 2009, \apj, 700, 684 

\bibitem[Mizuno et al.(2012)]{miz12} Mizuno, Y., Lyubarsky, Y., Nishikawa, K.-I., \& Hardee, P. E. 2012, \apj, 757, 16

\bibitem[Mizuno et al. (2008)]{miz08} Mizuno, Y., Hardee, P., Hartmann, D. H., Nishiakwa, K.-I., \& Zhang, B. 2008, \apj, 672, 72

\bibitem[Mizuno et al. (2015)]{miz15} Mizuno, Y., G\'{o}mez, J. L., Nishikawa, K.-I., Meli, A., Hardee, P. E. \& Rezzolla, L. 2015 \apj, 809, 38

\bibitem[Moll (2009)]{mol09} Moll, R. 2009, \aap, 507, 1203

\bibitem[Moll et al. (2008)]{mol08} Moll, R., Spruit, H. C., \& Obergaulinger, M. 2008, \aap, 492, 621

\bibitem[Mundell et al. (2013)]{mun13} Mundell, C. G., Kopac, D., Arnold, D. M., et al. 2013, \nat, 504, 119

\bibitem[Nagai et al. (2014)]{nag14} Nagai, H., Haga, T., Giovannini, G., et al. 2014, \apj, 785, 53

\bibitem[Nakamura et al. (2007)]{nak07} Nakamura, M., Li, H., \& Li, S. 2007, \apj, 656, 721

\bibitem[Nakamura \& Meier (2004)]{nak04} Nakamura, M., \& Meier, D. L. 2004, \apj, 617, 123

\bibitem[Narayan et al. (2009)]{nar09} Narayan, R., Li, J., \& Tchekhovskoy, A. 2009, \apj, 697, 1681

\bibitem[Nokhrina et al. (2015)]{nok15} Nokhrina, E. E., Beskin, V. S., Kovalev, Y. Y. \& Zheltoukhov A. A., 2015, \mnras, 447, 2726

\bibitem[O´Neill et al. (2012)]{one12} O'Neill, S. M., Beckwith, K., \& Begelman, M. C. 2012, \mnras, 422, 1436

\bibitem[Parker (1979)]{par79} Parker, E. N. 1979, Oxford, Clarendon Press; New York, Oxford University Press, 1979, 858 p.

\bibitem[Piotrovich et al. (2014)]{pio14} Piotrovich, M. Yu., Gnedin, Yu. N., Buliga, S. D., et al. 2014, arXiv:1409.2283

\bibitem[Porth (2013)]{por13} Porth, O. 2013, \mnras, 429, 2428

\bibitem[Porth \& Komissarov (2015)]{por15} Porth, O., \& Komissarov, S. 2015, \mnras, 452, 1089

\bibitem[Porth et al. (2014)]{por14} Porth, O., Xia, C., Hendrix, T., Moschou, S. P., \& Keppens, R. 2014, \apjs, 2014, 4

\bibitem[Pushkarev et al. (2009)]{pus09} Pushkarev, A. B., Kovalev, Y. Y., Lister, M. L., \& Savolainen, T. 2009, \aap, 507, L33

\bibitem[Rocha da Silva et al. (2015)]{roc15} Rocha da Silva, G., Falceta-Gon\c{c}alves, D., Kowal, G., \& de Gouveia Dal Pino, E. M. 2015, \mnras, 446, 104

\bibitem[Santos-Lima et al.(2010)]{2010ApJ...714..442S} Santos-Lima, R., Lazarian, A., de Gouveia Dal Pino, E.~M., \& Cho, J.\ 2010, \apj, 714, 442 

\bibitem[Santos-Lima et al.(2012)]{2012ApJ...747...21S} Santos-Lima, R., de Gouveia Dal Pino, E.~M., \& Lazarian, A.\ 2012, \apj, 747, 21 

\bibitem[Sikora et al. (2005)]{sik05} Sikora, M., Begelman, M. C., Madejski, G. M., \& Lasota, J.-P 2005, \apj, 625, 72

\bibitem[Singh et al.(2015)]{2015ApJ...799L..20S} Singh, C.~B., de Gouveia Dal Pino, E.~M., \& Kadowaki, L.~H.~S.\ 2015, \apjl, 799, L20 

\bibitem[Sironi  \& Spitkovsky(2014)]{2014ApJ...783L..21S} Sironi, L., \& Spitkovsky, A.\ 2014, \apjl, 783, L21 

\bibitem[Sironi et al. (2015)]{sir15} Sironi, L., Petropoulou, M., \& Giannios, D. 2015, \mnras, 450, 183

\bibitem[Takamoto et al. (2015)]{tak15} Takamoto, M., Inoue, T., \& Lazarian, A.\ 2015, \apj, 815, 16

\bibitem[Tomimatsu et al. (2001)]{tom01} Tomimatsu, A., Matsuoka, T., \& Takahashi, M. 2001, \prd, 64, 123003

\bibitem[Tchekhovskoy \& Bromberg(2015)]{Tch15} Tchekhovskoy, A., \& Bromberg, O. 2015, arXiv:1512.04526

\bibitem[Wiersema et al. (2014)]{wie14} Wiersema, K., Covino, S., Toma, K. 2014, \nat, 509, 201

\bibitem[Yuan \& Narayan (2014)]{yua14} Yuan, F., \& Narayan, R. 2014,  ARA\&A, 52, 529

\bibitem[Zhang \& Yan (2011)]{zha11} Zhang, B., \& Yan, H. 2011, \apj, 726, 90



\end{thebibliography}
\end{document}